\documentclass[12pt]{article}

\usepackage{amsmath,amssymb,epsfig}

\newcommand{\beq}{\begin{equation}}
\newcommand{\eeq}{\end{equation}}
\title{{\bf Neutrino mass matrix: inverted hierarchy and CP violation}}
\author{{Michele Frigerio} \footnote{\tt frigerio@he.sissa.it}\\
{\small\it  INFN, Section of Trieste {\rm and} International  School}\\ 
{\small\it for Advanced Studies (SISSA), Via Beirut
4, I-34014 Trieste, Italy.}\\
{Alexei Yu. Smirnov} \footnote{\tt smirnov@ictp.trieste.it}\\
{\small\it The Abdus Salam International  Center for Theoretical Physics
(ICTP), I-34100 Trieste,}\\
{\small {\it Italy} and \it Institute for  Nuclear Research, Russian Academy of
Sciences, Moscow, Russia.} }

\date{}

\begin{document}
\maketitle


\begin{abstract}
We reconstruct the neutrino mass matrix in flavor basis,
in the case of inverted mass hierarchy (ordering), using
all available experimental data on neutrino masses and  oscillations.
We analyze the dependence of
the matrix elements $m_{\alpha\beta}$ on the CP violating Dirac, $\delta$,
and Majorana, $\rho$ and $\sigma$, phases, for different values 
of the absolute mass
scale.
We find that the present data admit various structures of the mass matrix:
(i) hierarchical structures with a set of  small (zero) elements; (ii)
structures with equalities among various groups of elements:   
$e$-row and/or $\mu\tau$-block elements,
diagonal and/or off-diagonal elements; (iii) ``democratic''
structure. We find values of phases for which these structures are
realized. The mass matrix elements can anti-correlate
with flavor: inverted partial or complete flavor alignment is
possible. For various  structures of the mass matrix we identify
possible  underlying symmetry. We find that the mass matrix can be
reconstructed completely only in particular cases, provided that
the absolute scale of the mass is measured. Generally, the freedom related
to the Majorana phase $\sigma$  will not be removed,  thus
admitting various types of mass matrix.             

\vspace*{.3cm}
\noindent Ref.SISSA 57/2002/EP
\hfill
hep-ph/0207366
\end{abstract}
 
\begin{center}
PACS number: 14.60.Pq, 11.30.Er
\end{center}

\newpage

\section{Introduction}

Neutrino masses and mixing are considered to be the manifestation  of 
physics beyond the Standard Model. The question is: how far beyond? 
One way to 
answer is to confront various  models of neutrino
masses with experimental results in terms of 
mass squared differences and  mixing angles. 
In this approach  a typical situation 
is that predictive models (with
restricted number of free parameters) do not reproduce the data well. 
The  introduction of additional free  parameters 
allows to describe the data. However, in this case  the predictiveness is
lost.

In this connection,  it is worthwhile to elaborate on  the bottom-up 
approach: to reconstruct the underlying physics,  
or at least to get some hint of this physics, starting immediately from  
experimental data. The data include information on  
mass squared differences and mixing angles, which appear  as 
independent observables. The mass matrix  
unifies information on the masses and the angles, as well as on
possible complex phases,  and therefore may give
some additional insight. So, the bottom-up approach 
could consist of the following steps: 

(i) reconstruction of the mass matrix in the flavor basis, where the 
charged lepton mass matrix is diagonal; 

(ii) search for the symmetry basis and the energy  scale 
at which the underlying flavor symmetry could be  realized (broken); 
 
(iii) identification of  the underlying physics.  

Several remarks are in order: 

- complete determination of the mass matrix may 
be practically impossible; 

- it may happen that  there is no underlying symmetry at all;  

- neutrino mass matrix may receive several different contributions 
from different processes and mass scales. 

The hope is that the (at least partial) reconstruction  of the mass matrix in
flavor basis, the   searches of its  regularities and 
the study of dependence of the  matrix structure on basis may give
some hint of the mechanism of neutrino mass generation.\\ 
 
The neutrino mass matrices  compatible with neutrino
oscillation  data \cite{exp} have been extensively studied in
literature \cite{theo,altfer,massinv}.
However,  in most of previous works  (see, {\it e.g.}, \cite{altfer}) 
the structure of the mass matrix has been analyzed in the assumption of  
definite CP parities  of the three neutrinos, which is equivalent  to
absence of  CP violation. 
Moreover, exact bi-maximal mixing has been often considered.
Usual assumptions  are  either strongly hierarchical 
or completely degenerate spectra.
Special attention has been given to the small elements of the mass matrix.
In some recent works \cite{FGM} the 
matrices with two independent entries exactly equal to zero 
have been classified.

In the  paper \cite{MA}, we have analyzed in detail the structure 
of the Majorana mass matrix of neutrinos, 
for {\it normal} hierarchy (ordering) of the mass spectrum.
In this case, the electron flavor is concentrated in the two light
eigenstates.
We have found that the structure of the mass mass matrix strongly depends
on the CP-violating phases. New possible structures have been identified.
Parameterizations of the matrix in terms of powers of a
unique expansion parameter are given.

Here we will complete the analysis, studying the case 
of {\it inverted} hierarchy (ordering) \cite{altfer,massinv}.
In this case the electron flavor is mainly present in
the two heavy states.

In general, the spectra with normal and inverted hierarchy have
different phenomenology (cosmological consequences, absolute mass scale,
neutrinoless beta decay rate, oscillations). 
In oscillations the difference between the two spectra
appears if 1-3 mixing differs from zero. At present, the only observation
which could be sensitive to the mass hierarchy  is the neutrino burst
from SN1987A. It was shown that the data (especially, energy  spectra
detected by Kamiokande and IMB) can be better described in the case 
of normal hierarchy
with Earth matter effect to be taken into account \cite{SN}. The inverted
mass hierarchy is disfavored (see, however, \cite{barger}). 
These statements depend 
on the original neutrino spectra                                  
produced in  the star as well on the value of 1-3 mixing. Recent
calculations
show that the difference of the fluxes of different neutrino types 
can be rather small
\cite{newSN}, thus  diminishing possible oscillation effects
and therefore difference of predictions for normal
and inverted hierarchy.  

The goal of this paper is to present the most general study 
of possible structures of the mass matrix without 
additional assumptions. In particular, we perform comprehensive 
analysis  of  
dependence of the matrix elements on CP-violating 
phases. 
We will show that the assumptions of definite  CP-parities, bi-maximal
mixing, strictly hierarchical or degenerate spectrum 
and exactly zero elements exclude a number of interesting 
matrix structures.

The paper is organized as follows.
In section \ref{rec} we reconstruct the mass matrix in flavor basis 
and describe the method of our analysis.
In section \ref{stru} we study possible structures of the matrix:
hierarchical structures (section \ref{hiera}), structures with 
equalities of matrix elements
(\ref{equa}), structures with inverted flavor alignment (\ref{ordering}).
In section \ref{masss} we describe the dependence of the matrix structure
on the mass spectrum, analyzing
the case of strong inverted hierarchy (section \ref{sstrong}),
inverted ordering (\ref{IO}) and degeneracy (\ref{dege}).
In section \ref{conc} we summarize the main results of our analysis.

\section{Reconstruction of the mass matrix \label{rec}}

\subsection{Parameterization and experimental input \label{notation}}

We consider the mass and mixing pattern  for three Majorana 
neutrinos. 
The flavor neutrino states are related to the mass
eigenstates by the unitary  mixing matrix $U$:
\beq
\nu_{\alpha L}=U_{\alpha i}\nu_{iL}\;, \;\;\;\; \alpha =e,\mu,\tau\;,
\;\;\;\; i=1,2,3\;.
\nonumber
\end{equation}
In the flavor basis, the Majorana mass matrix  
 $M$ can be written as  
\begin{equation}
M = U^*M^{diag}U^{\dag}~,~~~~~~~~~~~
M^{diag} \equiv diag(m_1e^{-2i\rho},~m_2,~m_3e^{-2i\sigma}), 
\label{matr}
\end{equation}
where $m_i$ (i = 1, 2, 3) are the moduli of the neutrino mass eigenvalues;
$\rho$ and $\sigma$ are the CP violating Majorana phases,
varying between $0$ and $\pi$.

We use the standard parameterization for the mixing matrix $U$:
\begin{equation}
U =
\left( \begin{array}{ccc}
 c_{13}c_{12} & s_{12}c_{13} & s_{13} e^{-i\delta}
\\ -s_{12}c_{23}-s_{23}s_{13}c_{12} e^{i\delta} &
    c_{23}c_{12}-s_{23}s_{13}s_{12} e^{i\delta}
& s_{23}c_{13} \\ s_{23}s_{12}-s_{13}c_{23}c_{12} e^{i\delta}
 & -s_{23}c_{12}-s_{13}s_{12}c_{23} e^{i\delta} & c_{23}c_{13}
\end{array} \right)\;,
\label{U}
\end{equation}
where $c_{ij} \equiv \cos \theta_{ij}$, $s_{ij} \equiv \sin \theta_{ij}$
and $\delta$ is the CP violating Dirac phase. The mixing angles
vary between $0$ and $\pi/2$ and $\delta$ between $0$ and $2\pi$.

The choice of the  parameterization in Eqs.(\ref{matr},\ref{U}) is convenient 
once all elements of mass matrix 
are considered \cite{MA}.  In particular,  dependence of the matrix on the
phase  $\delta$ is associated with the small parameter $s_{13}$, so that  
the $\delta$-dependence disappears when $s_{13} \rightarrow 0$. 
Furthermore, in the case of strong mass hierarchy,  
$m_{3} \rightarrow 0$, the dependence of the matrix on the phase $\sigma$ 
also disappears.\\

We consider the mass and mixing pattern  which explains the atmospheric
neutrino results  by   $\nu_{\mu} - \nu_{\tau}$ oscillations as the
dominant mode  and solves  the solar neutrino problem via the LMA MSW
conversion.  Correspondingly,  the mass split between 
$\nu_1  (\nu_2)$ and  $\nu_3$ states 
is determined by  the atmospheric neutrino mass squared difference
$\Delta m^2_{atm}$ and the mass split between 
$\nu_1$ and  $\nu_2$ (``solar'' neutrino pair)  by the solar
mass squared difference $\Delta m^2_{sol}$.
In the case of {\it inverted}  mass hierarchy (ordering),    
the $\nu_1$ and $\nu_2$  states 
have masses larger than the third state   
and  $m_2$ is the largest mass ($m_2 > m_1 >  m_3$):   
\beq
m_1^2\equiv m_3^2+\Delta m^2_{atm}\;,\qquad 
m_2^2\equiv m_1^2+\Delta m^2_{sol}\; . 
\label{inv}
\end{equation}

We use the following experimental results from neutrino oscillations
\cite{exp},
given at  $90\%$ C.L.:
\beq
\begin{array}{l}
\Delta m^2_{sol} =
\left( 6.2^{~ +10}_{~ -3} \right) \cdot 10^{-5}~ {\rm eV}^2\;;\\
\Delta m^2_{atm}=
\left(2.5^{~ +1.4}_{~ -0.9} \right) \cdot 10^{-3} {\rm eV}^2\;;\\
\tan^2 \theta_{12}=  0.41^{ + 0.2}_{ - 0.1}\;;\\ 
\tan \theta_{23} = 1 ^{~+ 0.35}_{~- 0.25}\;;\\
s_{13} \lesssim 0.2\; .  
\end{array}   
\label{data}
\end{equation}
The absolute mass scale and the three CP violating phases 
will be considered  as free parameters.

Important restrictions on the possible structures of the mass matrix 
come from the 
{\it upper} bound  on the 1-2 mixing. The relevant quantity is the
deviation of 1-2 mixing from maximal, which can be characterized by 
$\cos 2\theta_{12}$. Recent analysis of the solar neutrino data
\cite{globalfits} gives: 
\beq 
\cos 2\theta_{12} > 0.25~~(0.16)~,~~90\% ~~{\rm C.L.}~~
(99\% ~~ {\rm C.L.})\;.
\label{upper12}
\end{equation}

It is convenient  to introduce
dimensionless parameters:
$$
r\equiv \dfrac{m_3}{m_2}\;,\quad ~~~~ k\equiv \dfrac{m_1}{m_2}\;.
$$
The hierarchy parameter, $r$, is given by: 
\beq
r \approx  \sqrt{1 -  \frac{\Delta m^2_{atm}}{m_2^2}}  
\label{re}
\end{equation}  
and, for strong inverted hierarchy (ordering), $r\ll 1$ ($r<1$).
A distinctive feature of the inverted mass spectrum is
that  the states $\nu_1$ and $\nu_2$ are strongly degenerate in mass,  
for any value of $r$, so that  $k$ is very close to 1: 
\beq
k \approx 1 - \epsilon, ~~~~~ 
\epsilon \equiv \frac{\Delta m^2_{sol}}{2(m_3^2+ \Delta m^2_{atm})}        
\leq \frac{\Delta m^2_{sol}}{2\Delta m^2_{atm}} \lesssim 10^{-2}\;.
\label{ke}
\end{equation} 
As a consequence, the solar mass scale turns out to be very weakly 
imprinted into the
structure of the mass matrix. \\

In general, the elements of the matrix are complex quantities: 
\beq 
{M}_{\alpha\beta} = {m}_{\alpha\beta} e^{i \phi_{\alpha\beta}}. 
\end{equation}
The absolute values of matrix elements, 
${m}_{\alpha\beta}$,  are physical parameters 
which  can, in principle, be directly measured  in experiment. 
In contrast, the phases $\phi_{\alpha\beta}$
are changed by renormalization of the 
wave functions of flavor neutrino states. 
As a  result,  
only three  linear combinations of  phases are independent and have
physical meaning. 
In what follows we will study,  mainly,  the absolute values of 
$M_{\alpha\beta}$, which give  straightforward information on the
matrix structure.

Due to  maximal or near maximal 2-3 mixing, in the analysis it is convenient
to divide the mass matrix elements in 
two groups: the $e$-row elements, $m_{ee}$, $m_{e\mu}$, $m_{e\tau}$,
and the $\mu\tau$-block elements, $m_{\mu \mu}$, $m_{\mu\tau}$, 
$m_{\tau\tau}$.

\subsection{The limit $s_{13}=0$ and $m_1=m_2$: zero order matrix 
\label{limit}}

Let us introduce the matrix for   $s_{13}=0$ and  
$\epsilon = 0$: 
\beq
m_{\alpha\beta}^0 \equiv m_{\alpha\beta}(s_{13}=0, \epsilon = 0)\;, 
\label{zeroapp}
\end{equation} 
which we will call the {\it zero order matrix}.    
This matrix  gives rather precise approximation and  
it  allows  one  to identify possible  dominant and sub-dominant 
structures of the  exact  matrix.  

It is useful to introduce
\beq
X \equiv xe^{i\phi_x} \equiv s_{12}^2  e^{-2i\rho} + c_{12}^2\;,
\end{equation}
where the absolute value, $x$, and the phase, $\phi_x$,  equal  
\beq
x=\sqrt{1-\sin^2{2\theta_{12}}\sin^2{\rho}}~,~~~~~~~~~ 
\phi_x = -\arctan\left(\frac{\sin{2\rho}}
{\cot^2\theta_{12}+\cos{2\rho}}\right)\;.
\label{xfx}
\end{equation}
The zero order matrix of moduli can be written as
\beq
{m}^0 = \sqrt{\frac{\Delta m_{atm}^2} {1 - r^2}} 
\left(
\begin{array}{ccc}
x & c_{23} \sqrt{1-x^2} & s_{23} \sqrt{1-x^2}\\
\dots & |c_{23}^2  x + s_{23}^2 r e^{-2i\sigma_x}| &
s_{23} c_{23} |- x + r e^{-2i\sigma_x}|\\
\dots & \dots & |s_{23}^2 x + c_{23}^2 r e^{-2i\sigma_x}|
\end{array}
\right)\;,
\label{matrix}
\end{equation}
where 
\beq
\sigma_x\equiv\sigma+\phi_x/2 
\label{sx}
\end{equation}
varies in the interval $0 \div \pi$.

Let us consider the properties of the matrix (\ref{matrix}). 

1)  It depends on four independent parameters:
$x=x(\rho,\theta_{12})$, $\sigma_x=\sigma_x(\sigma,\rho,\theta_{12})$, 
$r$, $\theta_{23}$. 
According to  the experimental input (\ref{data}), we find that these
parameters are restricted, at 90\% C.L., in the following ranges:
\beq
r\in [0,1)\;,\quad c_{23}^2\in [0.35,0.65]\;,\quad 
x\in [\cos2\theta_{12},1]=[0.25,1]\;,\quad \sigma_x\in [0,\pi)\;.
\label{range}
\end{equation}

2) Using the bounds (\ref{range}), we get  the following maximal and
minimal values
of the matrix elements:
\beq
{m}_{ee} \in m_2 [0.25,1]\;,~~~{m}_{e\mu(e\tau)}\in m_2 [0,0.6]\;,
~~~{m}_{\mu\mu(\mu\tau)(\tau\tau)}\in m_2 [0,1]\;.
\label{mami}
\end{equation}
For previous studied of the allowed values of the matrix elements see
\cite{rode}.

3)  CP is conserved only for extreme values of $x$: 
$$
\begin{array}{ll}
x = x_{min} \equiv \cos 2 \theta_{12}:  & \rho=\pi/2 \;; \\
x = x_{max} \equiv 1:  &  \rho=0\;.
\end{array}
$$

4) The best fit value of 1-2 mixing  (according to the  LMA solution of
the solar neutrino  problem) implies
\beq
x_{min}^{bf}  \approx 0.42 \;.
\label{xminbf}
\end{equation}
The upper bound (\ref{upper12}) on 1-2 mixing gives $x_{min} > 0.25$.  
These results  have important implications for the structure of the mass
matrix. In particular, $m_{ee}$ cannot be small: 
\beq
m_{ee} > \cos 2\theta_{12} m_2 > 0.25 m_2 > 0.25 \sqrt{\Delta m^2_{atm}}\;. 
\label{meelo}
\end{equation}
The $\mu\tau$-block elements in Eq.(\ref{matrix})
can be small only if $r$ is equal or larger 
than $x_{min}$, that is in the case of non hierarchical spectrum.

5)  The  six  elements of matrix (\ref{matrix}) 
are functions of only four parameters,
so that there are two relations among the elements:   
\beq
m_{ee}^2 + m_{e\mu}^2 +   m_{e\tau}^2 = m_2^2 \;,
\label{sumrel}
\end{equation}
\beq
(m_2^2 -  m_{ee}^2)(m_{\mu \mu}^2 - m_{\tau\tau}^2) = 
(m_{e\mu}^2 -  m_{e\tau}^2)(2 m_{ee}^2 - \Sigma_{\mu\tau})\;, 
\label{sumrel2}
\end{equation}  
where 
\beq
\Sigma_{\mu\tau} \equiv m_{\mu \mu}^2 + m_{\tau\tau}^2 + 2 m_{\mu\tau}^2
\label{sumrel3}
\end{equation} 
is the sum of $\mu\tau$-block elements squared. 
Moreover, the four  parameters are restricted to the ranges given in 
(\ref{range}). Therefore, the matrix structure is  constrained 
and there are correlations among the values of different elements. 

There are two other useful relations for the zero order matrix elements: 
\beq
m_{e\mu}^2 + m_{e\tau}^2 = (1 - x^2) m_2^2~,~~~~~
\Sigma_{\mu\tau} = (x^2+r^2) m_2^2~. 
\label{sumrel4}
\end{equation}
The sum of all matrix elements squared is equal to the sum of mass 
eigenvalues squared \cite{MA}:
\beq
\sum_{\alpha,\beta}m^2_{\alpha\beta}=(2+r^2)m_2^2\;.
\label{somma}
\end{equation}
This equality holds also when $s_{13}$-corrections are included.

\subsection{${\cal O}(s_{13})$ and  ${\cal O}(\epsilon)$ corrections 
\label{s13}}

The structure of the leading order (linear in $s_{13}$ and
$\epsilon$) corrections to $m^0$ can be parameterized by the matrices $m^s$    
and ${m}^{\epsilon}$: 
\beq
{m} = \left|{m}^0 + s_{13} {m}^{s}  + \epsilon {m}^{\epsilon} \right|
+ {\cal O}(s_{13}^2,\epsilon^2, s_{13} \epsilon )\;.
\label{0+s13}
\end{equation}

The upper bound on $s_{13}$ (see Eq.(\ref{data})) is not very strong   
and there is still room
for significant corrections to the matrix elements. 
Moreover, $s_{13}-$terms can give  dominant contribution 
if the elements in $m^0$ are small.

The main features of the  matrix  $m^s$ (its explicit expression is given
in the Appendix) are the following:

\begin{itemize}

\item ${m}_{ee}^{s}=0$, $m_{ee}$ receives  corrections only at the
order $s_{13}^2$. 

\item In the case of maximal atmospheric mixing ($\theta_{23}=\pi/4$), also
$m_{\mu\tau}^s = 0$. 
Moreover, corrections to $m_{e\mu}$ and $m_{e\tau}$ are equal and 
have opposite sign
(${m}^{s}_{e\mu} = - {m}^{s}_{e\tau}$). The same is true for $m_{\mu\mu}$
and  $m_{\tau\tau}$: ${m}^{s}_{\mu\mu} = - {m}^{s}_{\tau\tau}$.

\item The elements ${m}_{\alpha\beta}^{s}$ can be zero or take their maximal
value depending on the value of the Dirac phase $\delta$ 
(see Eq.(\ref{os13})). 

\end{itemize}
Thus, the ${\cal O}(s_{13})$ corrections change the splitting  
between $m_{e\mu}$ and $m_{e\tau}$ as well as  between
$m_{\mu\mu}$ and $m_{\tau\tau}$ elements.  
Moreover, this splitting depends strongly on $\delta$.

The smallness of $s_{13}$ could be a signal of an underlying symmetry.
The pattern of $s_{13}$ 
corrections to the mass matrix in flavor basis could suggest how
this symmetry is related to the flavor of neutrinos.

For maximal possible value of $s_{13}$,  
the terms $s_{13}{m}_{\alpha\beta}^{s}$ can be as large as  
$(0.1\div 0.2) m_2$.
Future experiments may strengthen  the upper bound on $s_{13}$, making the 
$s_{13}$ corrections even smaller.\\

Let us describe some general features of the $\epsilon$-corrections  
(the explicit expression  for ${m}^{\epsilon}$  is given in
the Appendix). 

\begin{itemize}

\item All matrix elements receive corrections proportional to $\Delta
m^2_{sol}$; $m^{\epsilon}_{\alpha\beta} = 0$ for particular values of
the phases $\rho$ and  $\sigma$ only.

\item In the case of maximal atmospheric mixing ($\theta_{23}=\pi/4$), 
corrections to $m_{e\mu}$ and $m_{e\tau}$ are equal to each other: 
${m}^{\epsilon}_{e\mu} = {m}^{\epsilon}_{e\tau}$.
The same is true for the corrections to $m_{\mu\mu}$ and  $m_{\tau\tau}$:
${m}^{\epsilon}_{\mu\mu}= {m}^{\epsilon}_{\tau\tau}$.

\end{itemize}

From Eq.(\ref{ke}) we get for the best fit values of 
$\Delta m^2$,  $\epsilon \lesssim  10^{-2}$,  
therefore the corrections ${m}^{\epsilon}_{\alpha\beta}$
are  about  $ 1 \%$ (using for
$\Delta m^2$ the ranges in (\ref{data}), we get $\epsilon\lesssim 0.05$). 
These corrections, however, have crucial 
phenomenological consequence: they break the degeneracy  
between $m_1$ and $m_2$  and thus explain the solar neutrino
conversion.  The value of the solar mass difference
emerges from minor details of the mass matrix (this is not the case for
normal hierarchy \cite{MA}).
The pattern of
${\cal O}(\epsilon)$ corrections could give some information about 
the origin of the small parameter $\Delta m^2_{sol}/\Delta m^2_{atm}$.

\subsection{$\rho-\sigma$ plots \label{rosigma}}

\begin{figure}
[p]
\begin{center}
\epsfig{file=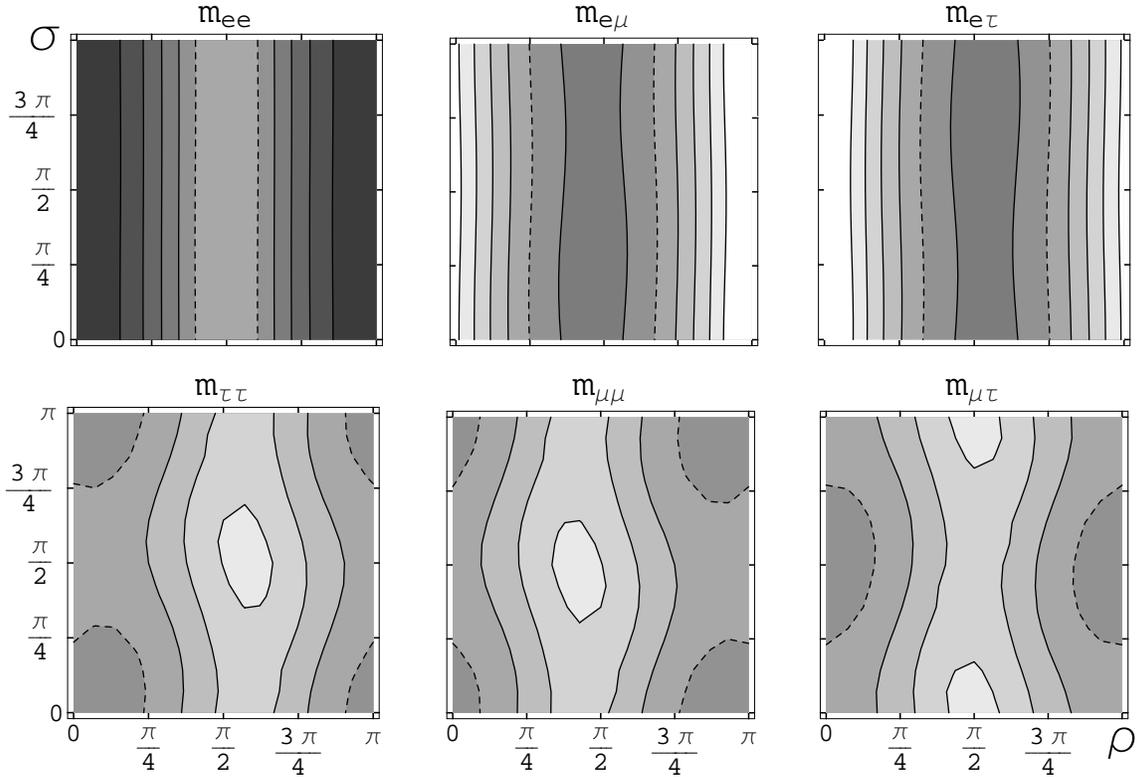,width=460pt}
\end{center}
\caption{The $\rho-\sigma$ plots for inverted hierarchical spectrum, 
with $r=0.1$. Contours are shown of constant mass (iso-mass) 
$m=(0.1,\,0.2,\,\dots,\,0.9)m^{max}$, where $m^{max}=0.05$ eV 
is the maximal value that the matrix elements can have, 
so that the white regions correspond to the mass interval 
($0-0.005$) eV and the darkest ones to ($0.045-0.05$) eV.
The contour $m = 0.5 m^{max}$ is dashed.  
We take $\Delta m^2_{sol}=6 \cdot 10^{-5} {\rm eV}^2$, 
$\Delta m^2_{atm}= 2.5 \cdot 10^{-3} {\rm eV}^2$ and 
$\tan^2\theta_{12}=0.4$, $\tan\theta_{23}=1$, $s_{13}=0.1$, 
$\delta=\pi/2$.}
\label{fig2}
\end{figure}

\begin{figure}
[p]
\begin{center}
\epsfig{file=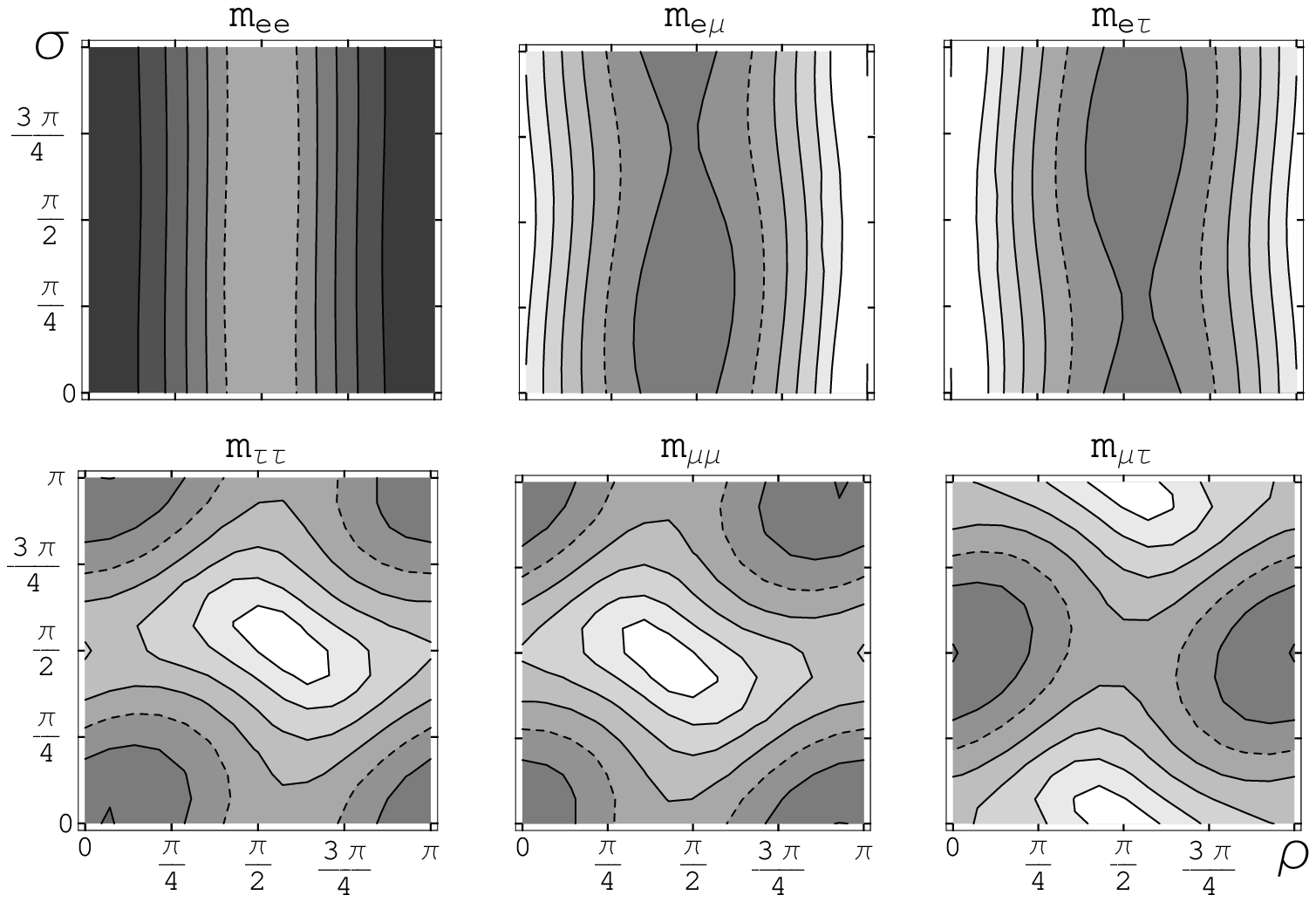,width=460pt}
\end{center}
\caption{The same as in Fig.\ref{fig2}, but for $r=0.4$.
In this case $m^{max}=0.055$ eV.}
\label{fig3}
\end{figure}

The dependence of 
the mass matrix on $s_{13}$ and, consequently, on the phase $\delta$, is 
rather weak and the  mass matrix is mainly determined by  
$x=x(\rho)$, $r$ and  $\sigma_x=\sigma_x(\rho,\sigma)$.  
Therefore,  to perform  a complete scanning of possible structures 
of the matrix, it is convenient to use the $\rho - \sigma$ 
plots which show lines of constant masses $m_{\alpha \beta}$ in the 
plane of Majorana phases $\rho$ and $\sigma$ \cite{MA}.  
In the Figs.\ref{fig2}-\ref{fig4}, we show the $\rho - \sigma$ plots for 
different values of the hierarchy parameter $r$. 

We have taken non-zero values for $\Delta m^2_{sol}$ and $s_{13}$, 
so that one can  identify their effects in the diagrams, as
deformations of the zero order form (\ref{matrix}) of the mass matrix.
For example, in Figs.\ref{fig2}-\ref{fig07delta} and in Fig.\ref{fig4}
we take $\theta_{23}=\pi/4$, which implies $m^0_{e\mu}=m^0_{e\tau}$
and $m^0_{\mu\mu}=m^0_{\tau\tau}$. The differences between the plots of 
these pairs of elements are due to $s_{13}$ terms.

Each point in the $\rho-\sigma$ plots \cite{MA} corresponds to physically
different mass matrix (obviously, the same point should be taken for 
all elements). The $\rho-\sigma$ plots allow one immediately to see 
(i) ranges in which a given matrix element can change, (ii) 
ranges of phases in  which a given element can be zero (small), 
(iii) correlations among values of different elements.

According to Figs.\ref{fig2}-\ref{fig4}, a 
large class of structures is allowed by the present data. 
Using  $\rho-\sigma$ plots, one can immediately identify the 
regions of parameters for which the matrix has: 

- hierarchical structure:  
 in which some elements are very small (white regions) and others are 
large (dark regions);

- non-hierarchical structure:  where all matrix elements 
are of the same  order -  have gray color; 
one can find structures with certain ordering of elements. 

- democratic structure: where all elements take the same or nearly the same 
value.

\section{Structures of the mass matrix \label{stru}}

In what follows, we will study possible structures of the mass matrix
using, first, the zero order approximation (\ref{matrix}) and, then,
evaluating the possible role of order $s_{13}$ corrections.

\subsection{Hierarchical structures and zeros 
\label{hiera}}

We will refer to the mass matrix structure as to a hierarchical one 
if some elements are smaller than others by a factor $0.1 \div 0.2$.  
Large elements (they should be of  order $m_2$) belong to the 
dominant structure, other elements to the sub-dominant structure.  
Also further structuring is possible within the dominant and sub-dominant
blocks. 
 
The hierarchical structures of the mass matrix may testify for the 
existence of certain symmetries.  

The hierarchical matrices can be found  by searching for zeroes
(small values) of some elements. They can be identified as white regions
in the $\rho-\sigma$ plots. In the analytical treatment, we use 
the zero order matrix (\ref{matrix}).

1) The element ${m}_{ee}$ cannot be zero (see Eq.(\ref{meelo})). 
It  belongs to the dominant structure of the matrix. 
The hierarchical structures with ${m}_{ee} \approx 0$, widely 
discussed in the literature \cite{altfer,massinv},  are strongly 
disfavored now. 

2)  The elements 
${m}_{e\mu}$ and ${m}_{e\tau}$ are simultaneously  zero 
(small) when $x=1$.  This corresponds to $\rho = 0$ 
and, consequently, $\phi_x=0$ and  $\sigma_x=\sigma$.
For $x=1$,  the zero order matrix (\ref{matrix}) becomes:
\beq
{m}^0= m_2\left(
\begin{array}{ccc}
1 & 0 & 0\\
\dots & |c_{23}^2  + s_{23}^2 r e^{-2i\sigma}| &
s_{23} c_{23} |- 1 + r e^{-2i\sigma}|\\
\dots & \dots & |s_{23}^2  + c_{23}^2 r e^{-2i\sigma}|
\end{array}
\right) \; . 
\label{erow0}
\end{equation}
The $s_{13}$  and 
$\epsilon$ terms  (section \ref{s13}) 
can give the main contributions to the elements 
${m}_{e\mu}$ and ${m}_{e\tau}$.

3) Using  Eq.(\ref{matrix}), we find conditions at which one 
of the elements of the $\mu \tau$-block is  zero: 
\beq
\begin{array}{cl}
{m}_{\mu\mu} = 0: & \cos{2\sigma_x} = -1,~~  
x = r\tan^2{\theta_{23}}\;;\\
{m}_{\tau\tau} = 0:  &  
\cos{2\sigma_x}=-1, ~~ r=x\tan^2{\theta_{23}}\;;\\
{m}_{\mu\tau} = 0:  &  
\cos{2\sigma_x} = 1, ~~ r=x\;.
\end{array}
\label{muta0}
\end{equation}
These analytic expressions describe the position of white regions in
the $\rho-\sigma$ plots, in  first approximation. 
Some additional shift of these regions 
is due to $s_{13}$ corrections (that is, zeros can be realized
for changed values of $\rho$ and $\sigma$); 
$\epsilon$ corrections are negligible. 

\begin{figure}
[p]
\begin{center}
\epsfig{file=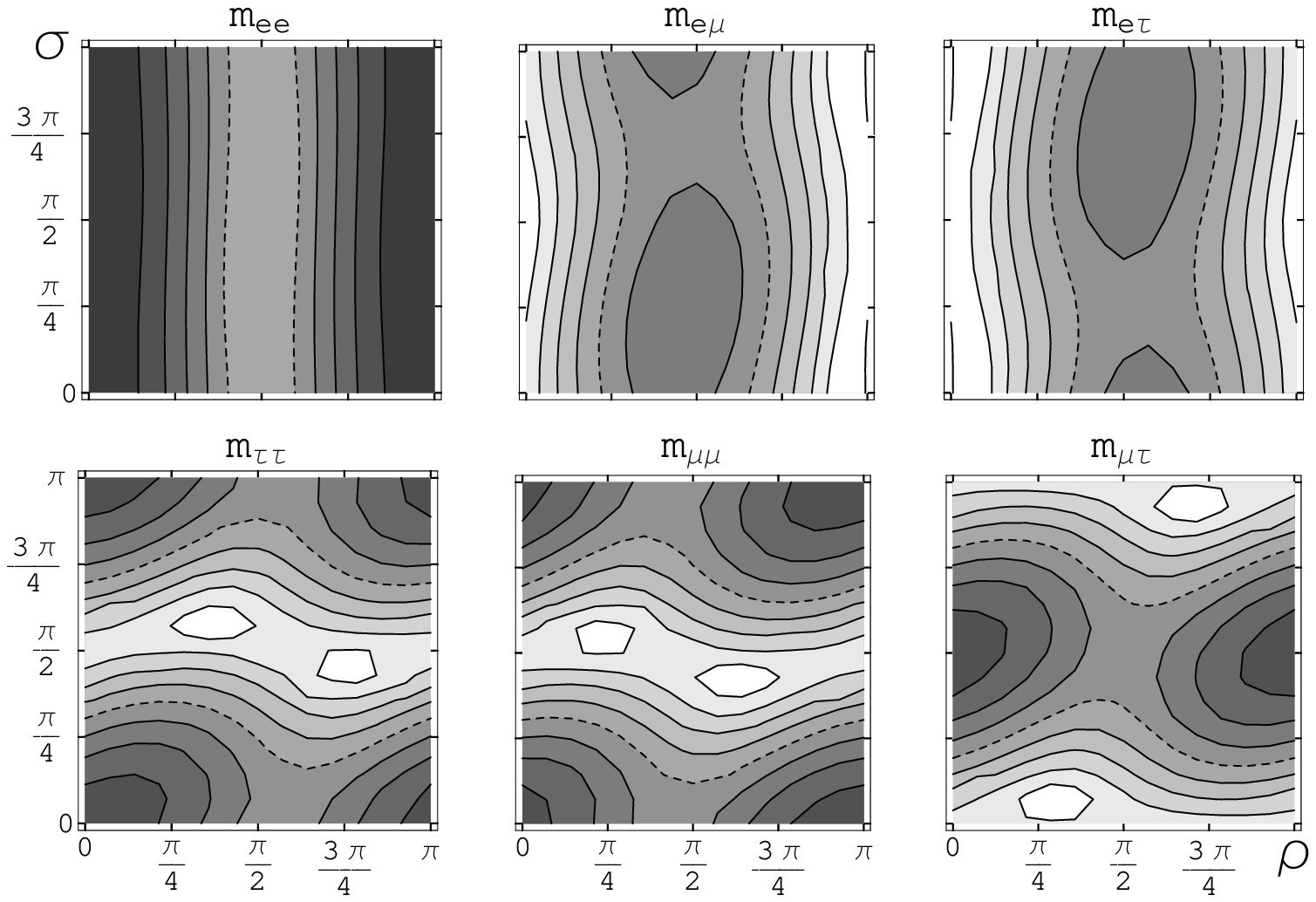,width=460pt}
\end{center}
\caption{The same as in Fig.\ref{fig2}, but for $r=0.7$.
In this case $m^{max}=0.07$ eV.}
\label{fig07}
\end{figure}

\begin{figure}
[p]
\begin{center}
\epsfig{file=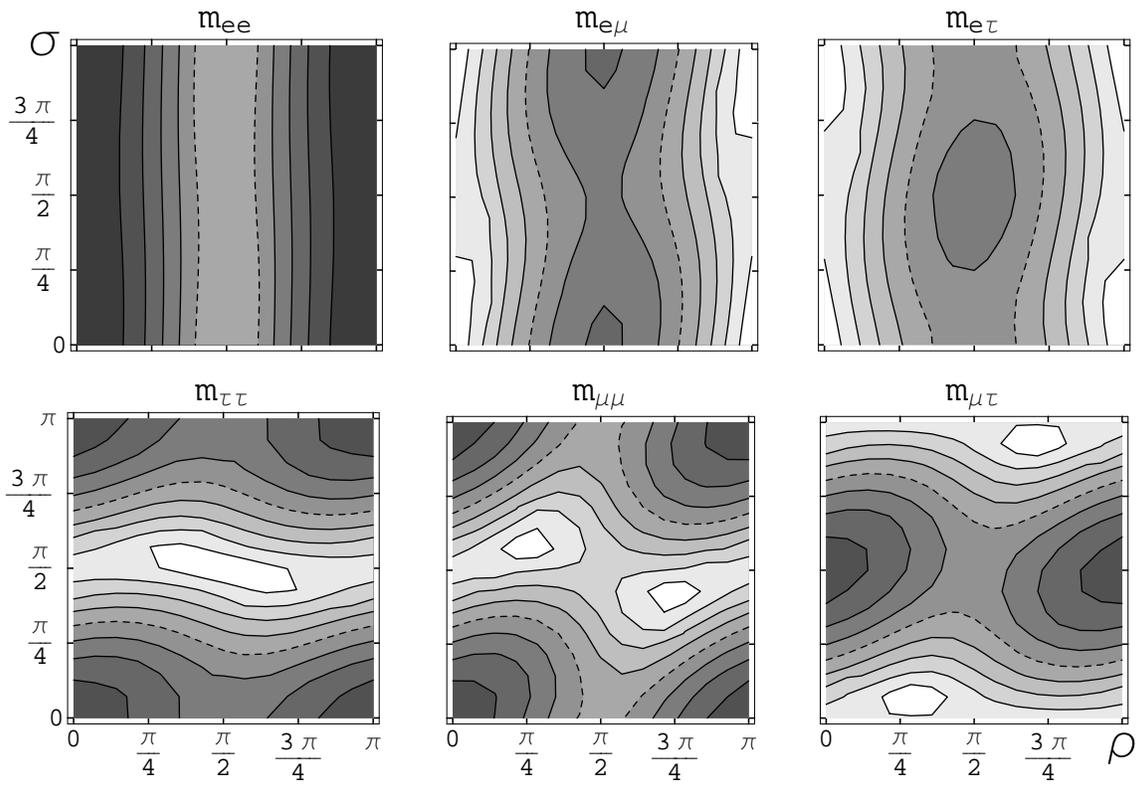,width=460pt}
\end{center}
\caption{The same as in Fig.\ref{fig2}, but for $r=0.7$ and $\delta=0$.}
\label{fig07delta}
\end{figure}

\begin{figure}
[b]
\begin{center}
\epsfig{file=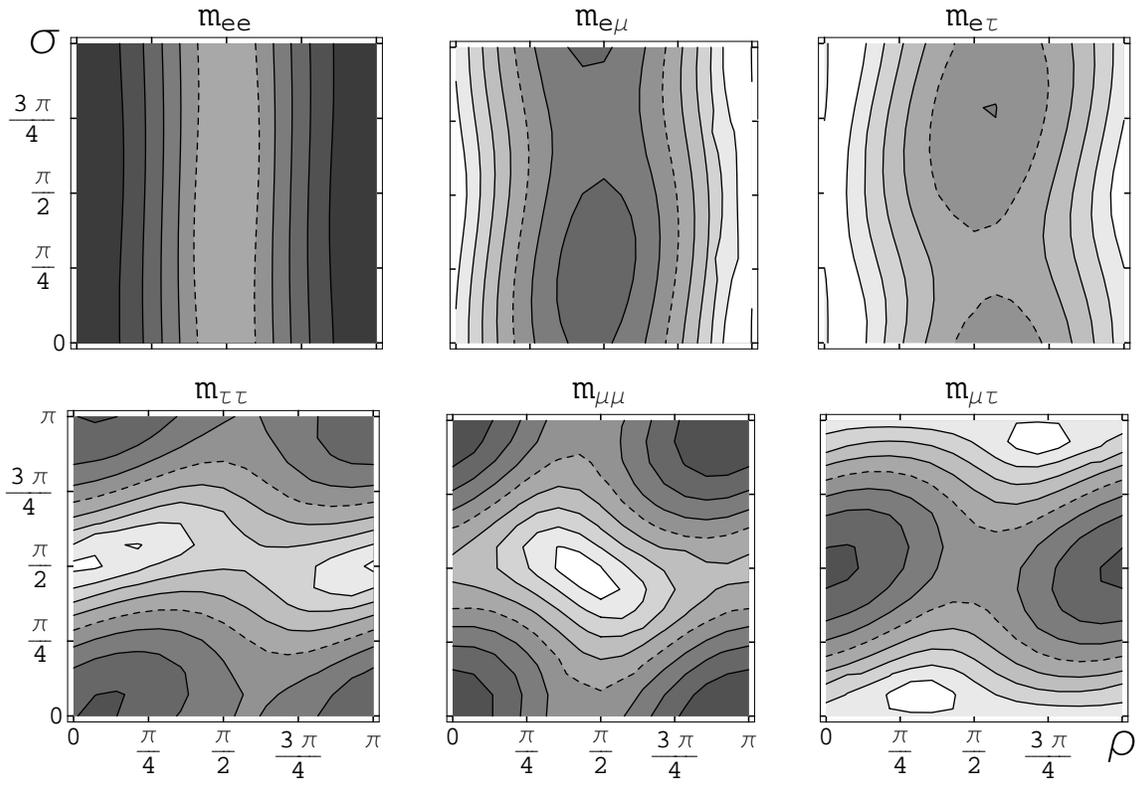,width=460pt}
\end{center}
\caption{The same as in Fig.\ref{fig2}, but for $r=0.7$ and 
$\tan\theta_{23}=0.75$.}
\label{fig07theta23}
\end{figure}

\begin{figure}
[p]
\begin{center}
\epsfig{file=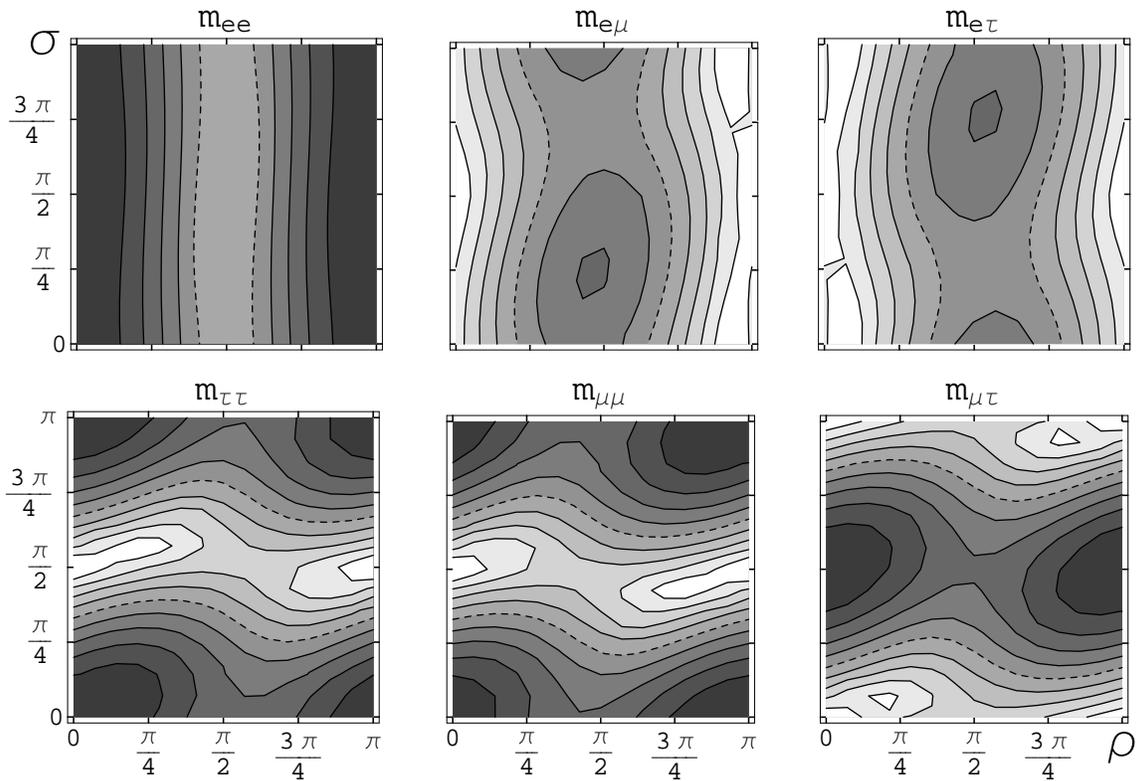,width=460pt}
\end{center}
\caption{The same as in Fig.\ref{fig2}, but for $r=0.99$: 
the spectrum is degenerate.
In this case $m^{max}=0.36$ eV, 
so that the white regions correspond to the mass interval 
($0-0.036$) eV and the darkest ones to ($0.324-0.36$) eV.} 
\label{fig4}
\end{figure}

As follows from (\ref{muta0}),   
almost maximal atmospheric mixing implies that 
the  $\mu\tau$-block elements can be zero only for $r\approx x$,
that is for {\it non-hierarchical} mass spectrum.  
The exact equality $x=r$ implies:
\beq
\sin^2{\rho}=\dfrac{1-r^2}{\sin^2{2\theta_{12}}}\;,
~~~~ r\ge \cos{2\theta_{12}}\;.
\label{minr}
\end{equation}
Therefore, only for $r$ large enough there is the possibility of 
hierarchical structures of the mass matrix with very small elements in the 
$\mu\tau$-block. 
Moreover, (\ref{minr}) shows that,
for increasing $r$, the regions of small $\mu\tau$-block elements move from
$\rho\approx \pi/2$ to $\rho\approx 0,\pi$, as one can see comparing
Figs. \ref{fig2}-\ref{fig4}.

Taking $\delta=0$ ($\pi$),  one can check that a white region appears
in the plot of $m_{\mu\mu}$ ($m_{\tau\tau}$) already for $r\approx 0.1$. 
This is a
case in which $s_{13}$ corrections are important:
deviation of $\theta_{23}$ from maximal value and relatively large
$s_{13}$ can add coherently,
increasing  the difference between $m_{\mu\mu}$
and $m_{\tau\tau}$.

In the case ${m}_{\mu\mu} = 0$, the matrix has the form 
\beq
{m}^0=
m_2 \left(
\begin{array}{ccc}
x & \sqrt{\frac{r(1-x^2)}{r+x}} & \sqrt{\frac{x(1-x^2)}{r+x}}\\
\dots & 0 & \sqrt{rx} \\
\dots & \dots & |x-r|
\end{array}
\right)\;, 
\label{mumu}
\end{equation}
where  $r = x \cot^2{\theta_{23}}$. Since  
$x - r = r(\tan^2{\theta_{23}} - 1)$,    
the element ${m}_{\tau\tau}$ is proportional to the deviation
of the atmospheric mixing from maximal one. 
In the case ${m}_{\tau\tau} = 0$, the matrix has an analogous form, 
but with the interchanges  $r \leftrightarrow x$
and ${m}_{\mu\mu}\leftrightarrow {m}_{\tau\tau}$.  
The structure (\ref{mumu}) is realized in Fig.\ref{fig07theta23},
for $\rho\approx\sigma\approx\pi/2$.

Both diagonal elements of the $\mu \tau$-block can be  zero 
at maximal 2-3 mixing  and  $x=r$,  so
that 
\beq
{m}^0= m_2 \left(
\begin{array}{ccc}
r & \sqrt{\frac{1-r^2}{2}} & \sqrt{\frac{1-r^2}{2}}\\
\dots & 0 & r \\
\dots & \dots & 0
\end{array}
\right)\;.
\label{mt0}
\end{equation}
This structure  is realized in Fig.\ref{fig3}, in the regions $\rho\approx
\sigma \approx \pi/2$ and in Fig.\ref{fig07}, for $\rho\approx\pi/4,3\pi/4$
and $\sigma\approx\pi/2$.

In the case ${m}_{\mu\tau} = 0$, the matrix (\ref{matrix}) 
has the form:
\beq
{m}^0= m_2 \left(
\begin{array}{ccc}
r & c_{23}\sqrt{1-r^2} & s_{23}\sqrt{1-r^2}\\
\dots & r & 0 \\
\dots & \dots & r
\end{array}
\right)\;.
\label{mutau}
\end{equation} 
Notice that the three diagonal elements are necessarily equal. 
This structure is shown in Fig.\ref{fig3}, for $\sigma \approx 0,\pi$
and $\rho\approx \pi/2$ and in Fig.\ref{fig07}, for $\sigma \approx 0$,
$\rho\approx \pi/4$ and also for $\sigma\approx\pi$,
$\rho\approx3\pi/4$.

4) The conditions for zero values of the $e$-row elements ($x=1$)  
and  $\mu \tau$-block elements are consistent with each other, so that one
may have any combination  of zeros in both blocks. 
In particular,  taking  $x=1$ in Eq.(\ref{mumu}), we get  
\beq
{m}^0= m_2\left(
\begin{array}{ccc}
1 & 0 & 0\\
\dots & 0 & \sqrt{r} \\
\dots & \dots & |1-r|
\end{array}
\right)\;,
\label{mumu0}
\end{equation}
or the same with $m_{\mu\mu}\leftrightarrow
m_{\tau\tau}$, if 
$m_{\tau\tau}=0$. This case is realized in Fig.\ref{fig07theta23},
for $\rho\approx0,\pi$ and $\sigma\approx\pi/2$.

If $r = 1$ (degenerate spectrum), in (\ref{mt0}) only ${m}_{ee}$ and 
${m}_{\mu\tau}$ differ from zero (and equal to $1$).
This hierarchical structure  is shown in Fig.\ref{fig4}, for
$\sigma \approx \pi/2$ and $\rho \approx 0,\pi$.

For $r=1$,  in  (\ref{mutau})  also $m_{e\mu}=m_{e\tau}=0$ and 
the matrix becomes the identity. 
This structure appears in Fig.\ref{fig4}, for 
$\rho\approx\sigma\approx 0,\pi$.\\

\begin{table}[p]
\begin{center}
\begin{tabular}{|c|c|c|c|c|}

\hline
& ${\cal O}(s_{13},\epsilon)$ entries & 
Range for $r$ & Range for $\tan\theta_{23}$ & $m^0/m_2$\\
\hline
\hline
I & $m_{e\mu}, m_{e\tau}$ & $0\le r\le1$ & $0.75 \div 1.35$ &
$
\left(
\begin{array}{ccc}
1 & 0 & 0\\
\dots & * & * \\
\dots & \dots & *
\end{array}
\right)
$
\\
\hline \hline
II & $m_{\mu\mu}$ & $0.2 \lesssim r \le 1$ & 
$\begin{array}{c} 0.75 \div 0.85 \\ 1.15 \div 1.35 \end{array}$ &
$
\left(
\begin{array}{ccc}
* & * & *\\
\dots & 0 & * \\
\dots & \dots & *
\end{array}
\right)
$\\
\hline
III & $m_{\tau\tau}$ & $0.2 \lesssim r \le 1$ & 
$\begin{array}{c} 0.75 \div 0.85 \\ 1.15 \div 1.35 \end{array}$ &
$
\left(
\begin{array}{ccc}
* & * & *\\
\dots & * & * \\
\dots & \dots & 0
\end{array}
\right)
$\\
\hline \hline
IV & $m_{\mu\mu}, m_{\tau\tau}$ &
$0.4 \lesssim r \lesssim 0.8$ & 
$0.95 \div 1.05$ &
$
\left(
\begin{array}{ccc}
* & * & *\\
\dots & 0 & * \\
\dots & \dots & 0
\end{array}
\right)
$\\
\hline
V & $m_{\mu\tau}$ &
$0.4 \lesssim r \lesssim 0.8$ & 
$0.75 \div 1.35$ &
$
\left(
\begin{array}{ccc}
* & * & *\\
\dots & * & 0 \\
\dots & \dots & *
\end{array}
\right)
$\\
\hline \hline
VI & $m_{e\mu}, m_{e\tau},  m_{\mu\mu}$ &
$0.6 \lesssim r \lesssim 0.8$ &
$1.15 \div 1.35$ & 
$
\left(
\begin{array}{ccc}
1 & 0 & 0 \\
\dots & 0 & * \\
\dots & \dots & *
\end{array}
\right)
$\\
\hline
VII & $m_{e\mu}, m_{e\tau}, m_{\tau\tau}$ &
$0.6 \lesssim r \lesssim 0.8$ & 
$0.75 \div 0.85$ &
$
\left(
\begin{array}{ccc}
1 & 0 & 0 \\
\dots & * & * \\
\dots & \dots & 0
\end{array}
\right)
$\\
\hline \hline
VIII & $m_{e\mu}, m_{e\tau},  m_{\mu\mu}, 
m_{\tau\tau}$ & $r \approx 1$ & 
$0.95 \div 1.05$ &
$
\left(
\begin{array}{ccc}
1 & 0 & 0 \\
\dots & 0 & 1 \\
\dots & \dots & 0
\end{array}
\right)
$\\
\hline
IX & $m_{e\mu}, m_{e\tau},  m_{\mu\tau}$ &
$r \approx 1$ & $0.75 \div 1.35$ & 
$
\left(
\begin{array}{ccc}
1 & 0 & 0\\
\dots & 1 & 0 \\
\dots & \dots & 1
\end{array}
\right)
$\\
\hline

\end{tabular}
\caption{Hierarchical structures of the mass matrix. The classification
is based on the very small (${\cal O}(s_{13},\epsilon)$) entries of 
the matrix, which are listed in the second column. 
In the third and fourth columns the corresponding allowed 
ranges for $r$ and $\tan\theta_{23}$ are given.
When $r\approx 1$, zero elements in $m^0$ receive also ${\cal O}(\eta)$
corrections (see Eq.(\ref{eta})).
The matrices ${m}^0$, shown  in the last
column, are  simplified forms of the structures given 
in Eqs.(\ref{erow0}, \ref{mumu}-\ref{mumu0}).
Parameters can be chosen in such a way  that
the elements denoted with
``$*$'' belong to the range $0.3-0.7$.
The ${\cal O}(s_{13})$ corrections can be as large as $0.2$ for
the elements $e\mu,~e\tau,~\mu\mu$ and $\tau\tau~$ (see Eq.(\ref{os13})).
\label{table}}
\end{center}    
\end{table}

The discussed   mass matrices with zero elements   
are shown in Table \ref{table}. 
In the Table we give also  the intervals of  $r$ and $\tan\theta_{23}$
for which the structures can be realized. 
These intervals are computed requiring non-zero elements to be quite large
($>0.3 m_2$), in order to clearly distinguish between dominant and 
sub-dominant blocks.
So,  the matrices  we have  found have 
hierarchical structure.

One can check, using analytic relations given in Eqs. (\ref{xfx} - \ref{sx}),
that all matrices with zeros can be obtained using 
$\rho,\sigma=0,\pi/2$, that is  definite CP-parities. 

The ``zero'' elements are  zeroes  up to
${\cal O}(s_{13},\epsilon)$ corrections.  
In general,  corrections  are small ($\sim 10\%$) and could be very small
($\sim 1\%$), if the upper bound on $s_{13}$  becomes more stringent. 
Moreover, in all the hierarchical structures with  
$m_{e\mu} =  m_{e\tau} = 0$ ($x = 1$), the $\mu \tau$-block elements have
no ${\cal O}(s_{13})$  corrections (see Eq.(\ref{os13})).
Notice that, when one takes the limit $r=1$, terms of order $\eta$ are 
neglected, where
\beq
\eta\equiv 1-r \approx \dfrac{\Delta m^2_{atm}}{2 m_2^2}\;
\label{eta}
\end{equation}
(see Eq.(\ref{re})). Therefore, in the case $r=1$, ``zero''
elements are  zeroes  up to
${\cal O}(s_{13},\eta)$ corrections.
For $m_2\gtrsim 0.2$ eV, one gets $\eta\lesssim 3\cdot 10^{-2}$.

In Fig.\ref{fig07} and Fig.\ref{fig07delta} different values of $\delta$
are used ($\pi/2$ and $0$). Therefore, the relative phase of zero order
matrix elements and ${\cal O}(s_{13})$ terms is different in the two figures.
One can see, in particular, how this changes the values of $\rho$ and $\sigma$
corresponding to very small matrix elements (white regions).

Even for  non-zero values of  $s_{13}$ and $\epsilon$ one may have 
exact zeros in the matrix. 
In the case of inverted 
mass spectrum, there are five possibilities of two (and only two) exact zero 
elements \cite{FGM}:

\begin{itemize}

\item The four cases $m_{e\tau}=m_{\mu\mu(\tau\tau)}=0$ or
$m_{e\mu}=m_{\mu\mu(\tau\tau)}=0$.
The elements $m_{e\mu}$ and $m_{e\tau}$ cannot be both zero exactly, 
otherwise there is no solar mixing.
However, our analysis shows that, if one is small, also the other is,
because they cannot be zero separately in the limit $s_{13}=\epsilon=0$.

\item The case $m_{\mu\mu}=m_{\tau\tau}=0$.
This possibility is present also in the limit
$s_{13}=\epsilon=0$ (Eq.(\ref{mt0})).

\end{itemize}

As far as study of possible matrix structures is concerned, 
the requirement of exact zero values of some elements can be misleading.
Indeed, some elements can be small or very small but non-zero.
The smallness of an element 
could be explained by some flavor symmetry. However, the
flavor symmetry is broken anyway: it is difficult to expect exact  zeros. 
Moreover, zeros which exist at tree level  
can be unstable under radiative corrections.

Notice that we have looked for zero (small) elements in flavor basis.
Symmetries could be realized in a different basis.
Exactly zero elements in symmetry basis 
will receive contributions from the diagonalization of charged lepton
mass matrix and of possible non-canonical kinetic terms.
Still small elements which appear in flavor basis can be relevant,
in particular if the
symmetry basis is close to the flavor basis.
An analogous remark is valid for the analysis of equalities 
among matrix elements,
which we study in the next section.

\subsection{Equalities of matrix elements \label{equa}}

Equalities of some  matrix elements can be considered as 
the signature of certain symmetry or certain  origin of the neutrino masses.  
\\

1) {\bf ``Democratic'' matrix} of moduli.  
The zero order matrix of moduli (\ref{matrix}) can have
all six equal elements,  
\beq
{m}^0= \frac{m_2}{\sqrt{3}} \left(
\begin{array}{ccc}
1 & 1 & 1\\
\dots & 1 & 1 \\
\dots & \dots & 1
\end{array}
\right)\; , 
\label{demo}
\end{equation}
if and only if 
\beq
\theta_{23}=\dfrac{\pi}{4}\,,\quad x = \sqrt{\dfrac 13}\,,\quad
r=1\,,\quad \cos{2\sigma_x}=0\,.
\label{parde}
\end{equation}
That is, the  ``democratic" matrix of moduli corresponds to  degenerate mass
spectrum, maximal 
2-3 mixing and large CP-violating phases. According to Eq.(\ref{xfx}),
the condition 
$x = \sqrt{1/3}$ gives 
\beq
\sin^2\rho = \frac{2}{3 \sin^2 2\theta_{12}}~.
\end{equation}
For the best fit value of 1-2 mixing we have $\sin\rho  \approx 0.9$.
Then, the condition $\cos{2\sigma_x}=0$ implies
(see  Eqs.(\ref{xfx},\ref{sx})) 
$\sin\sigma\approx 0.56$ or $\approx 0.83$.
The present solar neutrino data admit 
$\rho  = \pi/2$ ($\sigma=\pi/4,3\pi/4$). 
Non-zero $\rho$ and $\sigma$  lead to non-zero
phases $\phi_{\alpha\beta}$ of the matrix elements. 

Notice that the values of parameters (\ref{parde}) correspond to the structure 
(\ref{demo}) only for $s_{13}=0$.
Substituting (\ref{parde}) in Eq.(\ref{os13}) and taking
$s_{13}\approx 0.2$, we find that
$m_{e\mu}$ and $m_{e\tau}$ can receive corrections (with opposite sign) 
as large as
$35\%$ of their zero order value. Also $m_{\mu\mu}$
and $m_{\tau\tau}$ can be shifted by $30\%$ in opposite directions. 
The magnitude of 
corrections depends strongly on $\delta$.
In principle, for $s_{13}\ne 0$, parameters can be 
readjusted in order to recover the structure (\ref{demo}).
In particular, this requires a deviation from maximal 2-3 mixing.
If quite large $s_{13}$ and exactly maximal 2-3 mixing 
were found in future experiments, the democratic structure 
of moduli would be excluded.
\\

2) {\bf Equal $e$-row elements}.  
All the $e$-row elements in Eq.(\ref{matrix}) are  equal,  
${m}_{ee} = {m}_{e\mu} ={m}_{e\tau}$,
if and only if
\beq
\theta_{23}=\frac{\pi}{4}, ~~~  x = \sqrt{\frac 13} \;.
\label{ero}
\end{equation}
In this case, one has also ${m}_{\mu\mu} = {m}_{\tau\tau}$.
Under the condition (\ref{ero}), we get 
\beq
{m}^0 = \frac{m_2}{\sqrt{3}}\left(
\begin{array}{ccc}
1 & 1 & 1\\
\dots & \frac{1}{2} \left|1 + \sqrt{3} r e^{-2i\sigma_x} \right| &
\frac{1}{2} \left|1 - \sqrt{3} r e^{-2i\sigma_x}\right|  \\
\dots & \dots & \frac{1}{2} \left|1 + \sqrt{3} r e^{-2i\sigma_x} \right| 
\end{array}
\right).
\label{eroweq}
\end{equation}
The ratio $r$ is not restricted, so that equality of the $e$-row 
elements can be realized for any type of spectrum (hierarchical,
non-hierarchical, degenerate). 

Using the  free parameters $r$ and $\sigma_x$,  one can produce further
structures or  reach equalities  in the $\mu \tau-$block. 
If 
$$ 
\cos{2\sigma_x} = \pm \frac{\sqrt {3}(1 - r^2)}{2r}\;, 
$$
the diagonal elements (sign plus) or off-diagonal elements 
(sign minus) of the $\mu \tau$-block are equal to $e$-row elements 
and we arrive at the zero order matrices:  
\beq
\frac{m_2}{\sqrt{3}} \left(
\begin{array}{ccc}
1 & 1 & 1\\
\dots & 1 & a(r) \\
\dots & \dots & 1
\end{array}
\right), ~~~~~~~
\frac{m_2}{\sqrt{3}} \left(
\begin{array}{ccc}
1 & 1 & 1\\
\dots & a(r) & 1 \\
\dots & \dots & a(r)
\end{array}
\right)  , 
\label{4+2}
\end{equation}
where 
$$
a(r) \equiv \sqrt{\frac{3r^2 - 1}{2}}\;. 
$$
Notice that ${m}_{\mu\mu} = {m}_{\tau\tau} \propto a(r)$ in the first case and 
${m}_{\mu\tau} \propto a(r)$ in the second one 
can be zero, if $r^2=1/3$, 
but they can also be equal to
the other elements, as in (\ref{demo}), if $r=1$ (degenerate spectrum).

All  elements of the $\mu \tau-$block are equal to each other if  
$
r \cdot \cos 2\sigma_x = 0\;. 
$
This condition can be realized  for $r = 0$ 
(and arbitrary $\sigma_x$),  which corresponds to  strong 
mass hierarchy, or for $\sigma_x = \pi/4, 3\pi/4$ and arbitrary spectrum. 
In the latter case we get the matrix: 
\beq
\frac{m_2}{\sqrt{3}} \left(
\begin{array}{ccc}
1 & 1 & 1\\
\dots & b(r) & b(r) \\
\dots & \dots &  b(r)
\end{array}
\right)\;, ~~~~
\label{emte}
\end{equation}
where 
$$
b(r) = \frac{m(\mu\tau-{\rm block})}{m(e-{\rm row})}  
\equiv \frac{1}{2} \sqrt{3 r^2 +1} . 
$$
The ratio $b(r)$ changes from 1/2, for strong mass hierarchy,   to 
1,  for  degenerate spectrum.
The structure (\ref{emte}) suggests that the magnitude of matrix element
can be connected to the electron flavor.
\\

3) {\bf Equal $\mu\tau$-block elements}.
The $\mu\tau$-block elements can be equal, 
${m}_{\mu\mu} = {m}_{\mu\tau} = {m}_{\tau\tau}$,   
in three cases:

\begin{itemize}

\item[(a)] $\theta_{23}=\frac{\pi}{4}, ~~~ \cos{2\sigma_x}=0$.

The matrix has the following form: 
\beq
{m}^0 = m_2 \left(
\begin{array}{ccc}
x & \sqrt{\frac{1-x^2}{2}} & \sqrt{\frac{1-x^2}{2}}\\
\dots &  \frac 12 \sqrt{x^2+r^2} &  \frac 12 \sqrt{x^2+r^2}\\
\dots & \dots & \frac 12 \sqrt{x^2+r^2}
\end{array}
\right) \;.
\label{IHmax}
\end{equation}
There are several interesting particular cases. 
The $\mu\tau$-block  elements are equal to $m_{ee}$ if $r^2=3x^2$:
\beq
{m}^0 =
\frac{m_2}{\sqrt{3}}\left(
\begin{array}{ccc}
r & \sqrt{\frac{3-r^2}{2}} & \sqrt{\frac{3-r^2}{2}}\\
\dots & r & r \\
\dots & \dots & r
\end{array}
\right)\;.
\label{mtdiag}
\end{equation}
The elements $m_{e\mu}$ and $m_{e\tau}$
can be  larger than 
the other elements for relatively small $r$
(notice that $r > \sqrt{3} x_{min}$) or equal to them ($r=1$).
The case $r=0$, discussed in the literature \cite{altfer}, requires $x=0$
and therefore maximal solar mixing, which is now excluded by data.

If $r^2=2-3x^2$, 
the matrix (\ref{IHmax}) becomes:
\beq
{m}^0 =
m_2 \sqrt{\frac {1+r^2}{6}}\left(
\begin{array}{ccc}
\sqrt{\frac{2(2-r^2)}{1+r^2}} & 1 & 1\\
\dots & 1 & 1\\
\dots & \dots & 1
\end{array}
\right).
\label{1+5}
\end{equation}
All elements but ${m}_{ee}$ are equal.
For $r=0$, $m_{ee}$ is two times larger than the other elements;
for $r=1$, all the elements are equal.

\item[(b)]  $\theta_{23}=\frac{\pi}{4}, ~~~ r=0$.

This case corresponds to  strong inverted hierarchy. The zero order
mass matrix  can be immediately obtained from  (\ref{IHmax}) 
taking the limit $r \rightarrow 0$.  
The properties of this matrix will be discussed in detail in section 
\ref{sstrong}. 

\item[(c)]  
$x = r, ~~~ 2 \sin^2{\sigma_x} \sin^2{2\theta_{23}}=1$.

In this case the $\mu\tau$-block elements are all equal to
$m_3/\sqrt{2}$ and ${m}_{ee}=m_3$.
Also the element ${m}_{e\mu}$ (${m}_{e\tau}$) is equal to $m_3/\sqrt{2}$ 
if ${r^2}/{2(1-r^2)}=c_{23}^2$ ($=s_{23}^2$).
According to Eq.(\ref{range}), this implies $0.45 \lesssim r \lesssim 0.65$.
The mass matrix takes the form:
\beq
{m}^0 =
\frac{m_3}{\sqrt{2}}\left(
\begin{array}{ccc}
\sqrt{2} & 1 & \frac 1r \sqrt{2-3r^2}\\
\dots & 1 & 1 \\
\dots & \dots & 1
\end{array}
\right)\;
\end{equation}
(or the same with $m_{e\mu} \leftrightarrow m_{e\tau}$).

\end{itemize}

The equality of $\mu\tau$-block elements can be an indication of a 
flavor symmetry with the same charge assigned to $\nu_{\mu}$ and $\nu_{\tau}$,
such as $L_e-L_{\mu}-L_{\tau}$. This symmetry would imply also 
${m}_{e\mu} = {m}_{e\tau}$. 
Notice that this equality holds
in the cases (a) and (b).\\

4) {\bf Equal diagonal elements}.
There are two possibilities for ${m}_{ee} = {m}_{\mu\mu} =
{m}_{\tau\tau}$:

\begin{itemize}
\item[(a)] $\theta_{23} = \dfrac{\pi}{4}\;, ~~~~
\cos 2\sigma_x = \dfrac{3x^2 -  r^2}{2xr}\;$.

In this case, $x\le r\le 3x$. The mass matrix has the form
\beq
{m}=
m_2 \left(
\begin{array}{ccc}
x & \sqrt{\frac{1-x^2}{2}} & \sqrt{\frac{1-x^2}{2}}\\
\dots & x & \sqrt{\frac{r^2-x^2}{2}} \\
\dots & \dots & x
\end{array}
\right)\;.
\end{equation}
Imposing equalities of diagonal elements also with 
${m}_{e\mu}$ (with ${m}_{\mu\tau}$)
we reproduce the first matrix in Eq.(\ref{4+2}) (Eq.(\ref{mtdiag})).
Notice that the diagonal elements can be much larger than off-diagonal 
elements only in the limit $x \rightarrow 1$, $r \rightarrow 1$,  
which corresponds to  degenerate spectrum. 

\item[(b)] $x=r\;, ~~~~ \cos{2\sigma_x} = 1\;$.

The matrix is given by Eq.(\ref{mutau}).
If $r={c_{23}}/{\sqrt{1+c_{23}^2}}$ ($0.35\lesssim r\lesssim 0.65$), 
also ${m}_{e\mu}$ is equal to the diagonal elements and the matrix becomes
\beq
{m}=
m_3\left(
\begin{array}{ccc}
1 & 1 & \frac 1r \sqrt{1-2r^2}\\
\dots & 1 & 0 \\
\dots & \dots & 1
\end{array}
\right)\;.
\label{mutau2}
\end{equation}
The matrix with ${m}_{e\tau}$ equal to the
diagonal elements can be found by substituting $c_{23}\leftrightarrow
s_{23}$ and
${m}_{e\mu}\leftrightarrow {m}_{e\tau}$ in (\ref{mutau2}).\\

\end{itemize}

5) {\bf Equal off-diagonal elements}.  The conditions for the  equality 
${m}_{e\mu}={m}_{e\tau}= {m}_{\mu\tau}$ 
can be found from Eq.(\ref{matrix}): 
$$
\theta_{23}=\frac{\pi}{4}\;, ~~~~~
\cos 2\sigma_x = \frac{3x^2 +  r^2 - 2}{2xr}\;.
$$
In this case the mass matrix has the following form:
\beq
{m}^0 = \frac{m_2}{\sqrt{2}}
\left(
\begin{array}{ccc}
\sqrt{2}x & \sqrt{1-x^2} & \sqrt{1-x^2}\\
\dots & \sqrt{2x^2+r^2-1} & \sqrt{1-x^2} \\
\dots & \dots & \sqrt{2x^2+r^2-1}
\end{array}
\right)\;.
\label{heavy}
\end{equation}
Imposing equalities also with ${m}_{ee}$ (${m}_{\mu\mu}$), 
we get the second matrix in Eq.(\ref{4+2}) (Eq.(\ref{1+5})).

For $r=1$, we get, at the same time, equal diagonal and off-diagonal 
elements. In this case the parameter $x$ can vary between $1/3$ and $1$,
so that the ratio between off-diagonal and diagonal elements,
$\sqrt{\frac{1-x^2}{2x}}$, can vary between $2$ and $0$.
This kind of equality suggests a permutation symmetry $S_3$ of the flavor 
neutrinos \cite{FTY}.
\\

\subsection{Ordering structures and flavor alignment \label{ordering}}

As one can see in the $\rho-\sigma$ plane, there are regions where 
all the matrix elements are
of the same order (intermediate gray in the $\rho-\sigma$ plots).
In these regions 
the matrix may have certain  ``ordering'' structures. 

Do masses correlate with flavors? That is, are there any correlations 
between charged lepton and neutrino masses? 
We will call such a correlation the {\it flavor alignment}. 

One criterion 
of alignment (motivated by possible horizontal symmetry) 
can be introduced prescribing
different lepton charges, $q_{\alpha}$, for different flavor neutrino states, 
$\alpha = e, \mu, \tau$. Suppose that neutrino masses 
equal 
\beq
m_{\alpha \beta} = \lambda^{q_{\alpha} + q_{\beta}},~~~~~\lambda < 1. 
\end{equation}
Then   the alignment exists if $q_e >  q_{\mu} > q_{\tau}$.  The smaller 
$\lambda$ or/and the larger the difference  of charges $q_{\alpha}$, 
the stronger is the alignment. 
In the case  $q_{\mu} = q_{\tau}$ (which might be indicated by maximal 
$\mu - \tau$ mixing), one can speak of partial alignment,  associated
to the lepton number $L_e$. In this case all $\mu\tau$-block elements
are equal.

Let us consider first the possibility of partial alignment of  the zero
order matrix in the limit of maximal 2-3 mixing. Using the 
matrix (\ref{IHmax}), we find that $m_{ee} < m_{e\mu}$ provided that 
$x^2 < 1/3$ (which is consistent with Eq.(\ref{xminbf})). 
Then, the condition $m_{\tau \tau} > m_{e\mu}$  would require 
$r > 1$, which is impossible  for inverted mass spectrum. Thus, 
even partial alignment can not be achieved. At best one can get 
``democratic'' structure, $m_{ee} =  m_{e\mu} = m_{\tau \tau}$, 
if $x^2 = 1/3$ and $r = 1$.

If 2-3 mixing deviates from maximal, 
a split appears between  $m_{e\mu}$  and $m_{e\tau}$ as well as 
$\mu\tau$-block elements.  
However, the same consideration as above holds for averaged values of the
$e$-row and $\mu\tau$-block elements: $(m_{e\mu}^2 + m_{e\tau}^2)/2$ and 
$\Sigma_{\mu\tau}/4$ (see Eq.(\ref{sumrel4})).  
The $s_{13}$ corrections can split  $m_{\mu\mu}$ and $m_{\tau \tau}$ 
elements, but no improvement of alignment  can be obtained. 

In the case of inverted mass spectrum, {\it inverted alignment}, 
$q_e <  q_{\mu} < q_{\tau}$, is possible. 
The alignment can be partial ($q_e <  q_{\mu} = q_{\tau}$).
Taking maximal 2-3 mixing and $r = 0$ (which corresponds to maximal split 
of the elements), we find from (\ref{IHmax}):
$$ m_{ee} : m_{e\mu} : m_{\tau \tau} = 
1: \sqrt{(1 - x^2) / (2 x^2)} : \frac 12\;.
$$ 
The inequality 
$m_{ee} > m_{e\mu} > m_{\tau \tau}$   is satisfied for 
$1/3 < x^2 < 2/3$. In particular, one can get the matrix of the form
\beq
m^0=N\left(
\begin{array}{ccc}
1 & \lambda & \lambda \\
\dots & \lambda^2 & \lambda^2 \\
\dots & \dots & \lambda^2
\end{array}
\right) \;,
\label{fact}
\end{equation}
with $N/m_2=\lambda = \sqrt{1/2}$.

Also for $r\ne 0$ the matrix can be reduced to the form (\ref{fact}).
Taking $r^2=(1-2x^2)/x^2$ in (\ref{IHmax}) 
(this is possible if $1/3\le x^2\le 1/2$), 
one gets $N/m_2=x$ and $\lambda^2=(1-x^2)/(2x^2)$,
which in turn implies $1/2\le \lambda^2\le 1$.

A complete inverted alignment can be achieved for non-maximal 2-3 
mixing. 
Inserting  $r=0$, $x^2=1/2$ and $c_{23}^2=s_{23}$ into (\ref{matrix}),
we get
\beq
m^0=\frac{m_2}{\sqrt{2}} \left(
\begin{array}{ccc}
1 & \lambda & \lambda^2 \\
\dots & \lambda^2 & \lambda^3 \\
\dots & \dots & \lambda^4
\end{array}
\right) \;,
\label{2fact}
\end{equation}  
with $\lambda=\tan\theta_{23}\approx 0.79$.

The role of $\mu$ and $\tau$ flavors are interchanged if $c_{23}=s_{23}^2$.
In this case a structure analogous to (\ref{2fact}) is realized, with
$\lambda=\cot\theta_{23}\approx 0.79$.

Structures in which $\mu$ and $\tau$ flavors are associated 
with substantially  different mass scales are excluded. Indeed, the
difference $m_{\mu\mu}^2-m_{\tau\tau}^2$ is proportional to the small
parameter $\cos 2\theta_{23}$.
Moreover, if there is a strong ordering between 
$m_{\mu\mu}$ and $m_{\tau\tau}$, the element $m_{\mu\tau}$ is larger than
both of them (see Eqs.(\ref{mumu},\ref{mumu0})),
while flavor alignment would require an intermediate value.

Notice that free parameters $r$, $\rho$, $\sigma$, etc. 
can be found  for which no correlation of the masses and lepton charges 
of the matrix elements exist at all. This possibility can be called 
{\it flavor disorder}.

\section{Dependence of the matrix structure on the type of mass spectrum 
\label{masss}}

Let us analyze how the matrix structure depends on $r$.
We also consider perspectives to reconstruct the mass matrix in
flavor basis in future experiments. 

As follows from Eq.(\ref{matrix}),  in the limit $s_{13}=0,~k=1$, 
the $e$-row
elements do not depend on $r$, so only the structure of the
$\mu\tau$-block  depends  on the type of mass spectrum. 

We split our discussion in three
parts:  (i) strong inverted hierarchy: $r\approx 0$, practically 
$0 \leq r \lesssim 0.2$;
(ii) inverted ordering: $0.2 \lesssim r \lesssim 0.8$;
(iii) degeneracy: $r\approx 1$,  practically $0.8 \lesssim r \leq 1$.

\subsection{Strong inverted hierarchy \label{sstrong}}            

\begin{figure}[p]
\begin{center}
\epsfig{figure=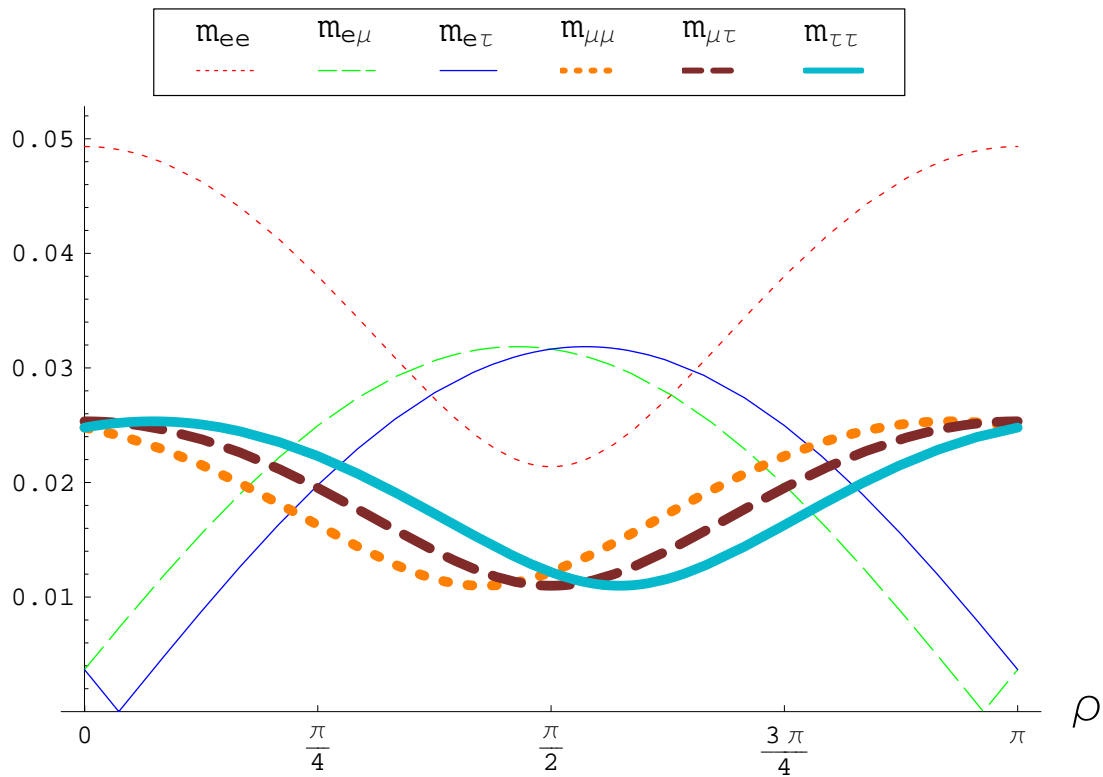,width=460pt}
\caption{Dependence of the absolute value of mass matrix elements (in eV) 
on $\rho$, in the case of mass spectrum with strong inverted hierarchy 
($r=0$). We take $\Delta m^2_{sol}=6 \cdot 10^{-5} {\rm eV}^2$, 
$\Delta m^2_{atm}= 2.5 \cdot 10^{-3} {\rm eV}^2$ and 
$\tan^2\theta_{12}=0.4$, $\tan\theta_{23}=1$, $s_{13}=0.1$, $\delta=\pi/2$.
\label{fig1}}
\end{center}
\end{figure}

Taking  $r=0$ ($m_3=0$) in (\ref{matrix}), 
we get the zero order matrix 
\beq
{m}^0= \sqrt{\Delta m^2_{atm}} x \left(
\begin{array}{ccc}
1 & c_{23} \sqrt{1/x^2 - 1} & s_{23}\sqrt{1/x^2 - 1}\\
\dots & c_{23}^2   &
s_{23} c_{23}  \\
\dots & \dots & s_{23}^2  
\end{array}
\right) 
\label{strong}
\end{equation}
which depends on two 
parameters only, $x$ and $\theta_{23}$. 
The dependence of $m^0$ on the Majorana phase $\sigma$, 
associated with $m_3$, disappears. Furthermore,
once $\theta_{12}$ and $\theta_{23}$ are fixed, the
matrix (\ref{strong}) depends only on $x=x(\rho)$.
In Fig.\ref{fig1},  we show
the absolute values of the matrix elements as functions of $\rho$.
The only freedom (associated to  
variations of  $\rho$ ($x$)) is reduced to change the relative size of
two groups of elements: $m_{ee}$ plus $\mu \tau$-block elements on  
one side and $m_{e\mu},  m_{e\tau}$ on  the other.

In the strong hierarchy case, we have (see Eq.(\ref{somma})): 
\beq
\sum_{\alpha,\beta} m_{\alpha \beta}^2 = 2 m_2^2=2\Delta m^2_{atm}\; . 
\end{equation} 
The sums of  $e-$row and $\mu \tau -$block elements are 
(see Eq.(\ref{sumrel4})): 
\beq
m_{ee}^2 + 2(m_{e\mu}^2+m_{e\tau}^2)  = m_2^2 (2 - x^2)\;,~~~~~~ 
\Sigma_{\mu\tau} = m_2^2 x^2=m_{ee}^2\;. 
\label{sumru}
\end{equation}  
So, the $e$-row elements dominate
over the $\mu \tau -$block elements. 
They are comparable  for $x = 1$, which corresponds to $m_{ee} = m_2$.  
The second equality in (\ref{sumru}) quantifies  
the dominance of the $ee-$element.

The mass matrix could be determined completely if direct measurements 
of neutrino masses were
sensitive to $m \approx \sqrt{\Delta m^2_{atm}}$. This can be checked
by future cosmological measurements, 
which will be  sensitive to sum of neutrino 
masses of the order of 0.1 eV \cite{astro}. 
Then, $x$ parameter can be found if 
$m_{ee}$ is measured in the neutrinoless double beta decay. 

In the matrix (\ref{strong})
$m_{ee} = \sqrt{\Delta m^2_{atm}} x$ is non-zero.  
Furthermore, the $\mu\tau$-block elements vary
in restricted ranges:
$$
m^0_{\mu\mu,\tau\tau}\in  m_2[0.1,0.65]\;,~~~~~~
m^0_{\mu\tau}\in m_2 [0.1,0.5]\; , 
$$
so that they cannot be zero either.  
Therefore, the only hierarchical structure which appears 
in the case of inverted hierarchy corresponds to 
$m_{e\mu} \approx m_{e\tau} \approx 0$.   These masses are small,
simultaneously, for $x\approx 1$ ($\rho\approx 0,\pi$). 
Substituting $x=1$ in Eq.(\ref{strong}), we get
\beq
{m}^0= \Delta m^2_{atm} \left(
\begin{array}{ccc}
1 & 0 & 0\\
\dots & c_{23}^2  & s_{23} c_{23}  \\
\dots & \dots & s_{23}^2 
\end{array}
\right) \;,
\label{hie}
\end{equation}
which is a particular case of the hierarchical structure (\ref{erow0}). 
In the limit $s_{13}=m_3=0$ and $x=1$, CP violation is absent ($\delta$ 
and $\sigma$ are irrelevant and $\rho=0$).

The hierarchical structure with $m_{ee} \approx m(\mu\tau-block) \approx 0$,
widely
discussed  in literature \cite{altfer,massinv} 
in connection to  $L_e-L_{\mu}-L_{\tau}$ symmetry, 
is strongly disfavored now. 
For the allowed values of $x$, the symmetry has to be strongly broken,
or realized in a  basis which differs from the flavor one \cite{mong}.

For maximal atmospheric mixing, the mass matrix (\ref{strong}) takes the form:
\beq
m^0= \sqrt{\Delta m^2_{atm}} x \left(
\begin{array}{ccc}
1 & \sqrt{\frac{1-x^2}{2x^2}} & \sqrt{\frac{1-x^2}{2x^2}}\\
\dots & 1/2 & 1/2 \\
\dots & \dots & 1/2
\end{array}
\right) \;.
\label{IH}
\end{equation}
Depending on $x$, the ratio between $m_{e\mu}$ ($m_{e\tau}$) 
and the other matrix elements can strongly change, as shown in
Fig.\ref{fig1}. 
Three interesting cases,
\beq
\left(
\begin{array}{ccc}
1 & 0 & 0\\
\dots & 1/2 & 1/2 \\
\dots & \dots & 1/2
\end{array}
\right) \;,\qquad
\sqrt{\frac 23}\left(
\begin{array}{ccc}
1 & 1/2 & 1/2\\
\dots & 1/2 & 1/2 \\
\dots & \dots & 1/2
\end{array}
\right) \;,\qquad
\sqrt{\frac 13}\left(
\begin{array}{ccc}
1 & 1 & 1\\
\dots & 1/2 & 1/2 \\
\dots & \dots & 1/2
\end{array}
\right) \;,
\label{3m}
\end{equation}
are  realized for $x=1$, $x=\sqrt{2/3}$ and 
$x=\sqrt{1/3}$, respectively. 
Only the first of these three structures, 
which corresponds to CP conservation ($\rho=0$),
has been considered before~\cite{altfer}.  
It is a particular case of (\ref{hie}). The second and the third matrices 
are particular cases of (\ref{1+5}) and (\ref{emte}), respectively.

As follows from  Eq.(\ref{os13}), for $r=0$ and $\theta_{23}=\pi/4$, 
the  ${\cal O}(s_{13})$ corrections to the elements of the matrix (\ref{IH}) 
have very simple form: 
$$
\begin{array}{l}
{m}^{s}_{ee} = {m}^{s}_{\mu\tau} = 0 \;,\\
{m}^{s}_{e\tau} = - {m}^{s}_{e\mu} = m_2  x \cos\varphi_1 /\sqrt{2} \;,\\
{m}^{s}_{\tau\tau} = - {m}^{s}_{\mu\mu} =   
m_2 \sqrt{1-x^2}\cos(\varphi_2-\phi_x)\;,
\end{array}
$$
where $\varphi_1$ and $\varphi_2$ are defined in Appendix.
Taking into account these corrections, one can explain the details of
Fig.\ref{fig1}.
In particular, for $\rho=0$ ($x = 1$, first matrix in Eq.(\ref{3m})),
the corrections  to the $\mu\tau$-block 
elements disappear (${m}^{s}_{\tau\tau}=0$), 
and  $s_{13}-$terms give dominant 
contributions to the $e$-row  elements   
(${m}^{s}_{e\tau}=  m_2 \cos\delta /\sqrt{2}$).  

When $m_{e\mu}^0\approx m_{e\tau}^0\approx 0$ ($x \approx 1$),
${\cal O}(s_{13})$ corrections can be used to get 
the inequality $m_{e\mu}\gg m_{e\tau}$
(or {\it vice versa}). Indeed, introducing the small parameter
$\gamma\equiv \sqrt{1-x^2}$, we find:
$$
m_{e\mu}\approx \frac {m_2}{\sqrt{2}}\left|\gamma-s_{13}\sqrt{1-\gamma^2}
\cos\varphi_1\right|\;,~~~~~
m_{e\tau}\approx \frac {m_2}{\sqrt{2}}\left|\gamma+s_{13}\sqrt{1-\gamma^2}
\cos\varphi_1\right|\;.
$$
Choosing $\delta$ such that $\cos\varphi_1=-1$, one gets 
$m_{e\mu}\gg m_{e\tau}$ for $\gamma\approx s_{13}\sqrt{1-\gamma^2}$.

\subsection{Inverted ordering \label{IO}}

This is a rather generic case, in which almost all structures 
discussed in section \ref{stru} can appear. 
In particular, all hierarchical structures but  VIII and IX 
can be realized (see Table \ref{table}). 
One or two 
$\mu\tau$-block elements  can be zero. 
As far as equalities of matrix elements are concerned, only the 
exact ``democratic'' structure is excluded. 
On the other hand, only in the inverted ordering case certain correlations of
masses and lepton charges appear. 
 
For a small value of $r$, the $\mu\tau$-block elements cannot be
very small and their dependence on $\sigma$ 
is weak, therefore the unique possible
hierarchical structure is I (Eq.(\ref{erow0})),
as one can see in Fig.\ref{fig2} ($r=0.1$).

For larger values of $r$,
the modifications in the $\mu\tau$-block elements can be strong.
We have seen, in section \ref{hiera}, that $\mu\tau$-block elements can be
very small only for $x\approx r$. 
According to Eq.(\ref{muta0}),
one can get $m_{\mu\mu}$ or
$m_{\tau\tau}$ equal
to zero for values of $r$ as small as $\sim x_{min}/2$,
because of  non-maximal 2-3 mixing.
Therefore, the structures II and III can be realized for $r\gtrsim 0.2$.
Instead, the equalities  $m_{\mu\tau}=0$ or
$m_{\mu\mu}=m_{\tau\tau}=0$ 
can be realized only for $r$ as large as $\sim x_{min}\gtrsim 0.4$ 
(structures IV and V).

If one requires that $m_{e\mu}$ and $m_{e\tau}$ are small together with
some $\mu\tau$-block elements, the condition $x\approx 1$ enforces
the minimal value of $r$ to be larger: $r_{min}\sim 1/2$ for the structures 
VI and VII; $r_{min}\sim 1$ for VIII and IX.
These considerations lead to the lower bounds
for $r$ given in the third column of Table \ref{table}.\\

In the non-hierarchical case 
$m_2 > \sqrt{\Delta m_{atm}^2}$. 
The absolute mass scale increases with $r$, {\it e.g.}, for $r \sim
0.8$ we get $m_2 \sim  0.1$ eV and
$m_{ee} \sim (0.03 \div 0.1)$ eV. 
Measuring $m_2$ and $m_{ee}$ we can immediately determine  
$x$, provided that $s_{13}$ is further restricted by experiment. 
Then $m_2$ and $\Delta m_{atm}^2$ determine $r$ according to 
(\ref{re}). The  only unknown parameter in the (zero order) matrix 
will be  $\sigma$. Its variations can strongly change the structure of the 
$\mu\tau$-block.

\subsection{Degenerate spectrum \label{dege}}

In the case of degenerate spectrum, the mass matrix coincides practically
with that for normal mass ordering \cite{MA}. The information about the
type of mass ordering  is imprinted in small,  ${\cal O}(\Delta
m^2_{atm}/m^2_2)$, deviation of $r$ from 1 (for inverted ordering $r <
1$).

A number of various mass structures are allowed in this case. 
The hierarchical structures I - III as well as VIII, IX can 
be realized. Moreover, the matrices VIII and IX appear only in the degenerate
case, taking the limit $r \rightarrow 1$ 
in Eq.(\ref{mt0}) and Eq.(\ref{mutau}),
respectively.
When the matrix approaches the  identity (IX), also 
the atmospheric mixing $\theta_{23}$ is generated by the small
corrections to the dominant structure.

The structures IV and V can not be realized: they require 
$r = x$ and, consequently,  $x = 1$. In turn, the latter implies 
$m^0_{e\mu} = m^0_{e\tau} = 0$. Also the structures VI and VII are forbidden
if $r\approx 1$, because they require $x\approx 1$: $x=r=1$ implies 
$m^0_{\mu \mu} = m^0_{\tau \tau}$.
\\

In the limit $r\rightarrow 1$, the zero order mass matrix 
depends on two unknown parameters: $x$ and $\sigma_x$. 
The first one can be determined directly from kinematic measurements 
of the absolute mass scale and detection of the neutrinoless $2\beta$ decay: 
$x \approx m_{ee}/m_2$. The only free parameter to which we have no access is 
the phase $\sigma$. Dependence of the $\mu \tau$-block elements on this phase 
is even stronger than in the inverted ordering case.

\section{Discussion and conclusions \label{conc}}

We have analyzed the structure of the Majorana mass matrix of the three
flavor neutrinos, in the case of inverted mass hierarchy (ordering). 

The structure of the mass matrix strongly depends on the Majorana phases 
$\rho$ and $\sigma$. 
We find that $e$-row elements strongly depend on $\rho$ and 
very weakly on $\sigma$. In contrast, the $\mu\tau$-block elements 
depend both on $\rho$ and $\sigma$, moreover the dependence on 
$\sigma$ becomes stronger with increase
of degeneracy. The dependence of the matrix on the Dirac phase 
$\delta$ is weak, because it is associated with the small parameter $s_{13}$.\\

The dominant structures of the mass matrix are determined, 
essentially, by four parameters:
$r= {m_3}/{m_2}$,
$\theta_{23}$,
$x=x(\rho,\theta_{12})$
and $\sigma$.

We find that present data allow for a large variety of different
mass matrix structures.

1) The hierarchical structures have a set of small or zero elements and 
a set of large elements. Any element but $m_{ee}$ can
be zero. The elements 
$m_{e\mu}$ and $m_{e\tau}$ can be very small for any type of
the mass spectrum. In addition,  one or two elements of the 
$\mu\tau$-block can be very small for inverted ordering  
and degenerate spectra. 

All the hierarchical structures 
can be realized for definite CP-parities of the mass eigenstates. 
The structures II,III and VI,VII (see Table \ref{table}) 
are allowed only if there is a deviation
from maximal atmospheric mixing.
In the case of strong inverted hierarchy, only the structure I is possible.
The structures IV-VII are allowed only for inverted ordering spectrum,
while VIII and IX only in the case of degeneracy.

The unique hierarchical structure which is present in the whole
range $0\le r\le 1$ is I (in fact it is present also for normal ordering of
the mass spectrum, $1\le r\lesssim 3$, but not for normal hierarchy, $r>3$).
This means that this structure is stable under modifications of the 
neutrino mass scale (in fact also under inversion of the  ordering).

2) Various equalities between  matrix elements are possible.
In particular, equalities of the $e$-row elements or/and 
$\mu\tau$-block elements, or diagonal elements or/and off-diagonal elements 
can be achieved. The ``democratic'' mass matrix is
also allowed in the flavor basis. 
Some equalities of elements (in contrast to zeros) can be realized
for non-trivial phases only.

3) We have studied correlations between masses and flavors. We 
find that flavor alignment is  impossible. However, one can reach 
inverted flavor alignment for rather large values of the expansion 
(ordering) parameter: $\lambda = 0.5 - 0.8$. 
Also flavor ``disorder'' is not excluded. 

4) We have shown that
${\cal O}(s_{13})$ 
and ${\cal O}(\Delta m^2_{sol}/\Delta m^2_{atm})$ terms 
can be as large as $(0.1\div 0.2)m_2$ and $(0.01\div 0.02)m_2$,
respectively.
Terms proportional to
$s_{13}$ depend on the CP violating Dirac phase $\delta$.
The values of $s_{13}$, $\delta$ and $\Delta m^2_{sol}$ are related to
small details of the matrix structure. 
In the degenerate case, also $\Delta m_{atm}^2$ is very weakly 
imprinted in the structure of the mass matrix.

One interesting possibility, proposed in a recent paper \cite{joshi},
is to generate radiatively
$s_{13}$ and $\Delta m^2_{sol}$, starting from a leading order matrix 
at high energy in which they are zero.

If $s_{13}$ stays at the present upper bound ($\sim 0.2$), ${\cal O}(s_{13})$
corrections can modify significantly the matrix structure, because they
shift in opposite directions the elements $m_{e\mu}$ and $m_{e\tau}$,
$m_{\mu\mu}$ and $m_{\tau\tau}$.\\

In general, the normal hierarchy spectrum \cite{MA} corresponds 
to a mass matrix with dominant $\mu\tau$-block. Flavor alignment is possible.
{\it Vice versa}, in the case of inverted hierarchy, there is dominance
of the $e$-row elements or, at least, of the $ee$-element. Inverted flavor 
alignment is possible. If the absolute mass scale increases, the spectrum 
becomes closer to the degenerate one and the difference between matrices 
which correspond to normal and inverted spectrum practically disappears.
The democratic mass matrix of moduli is possible only 
for exactly degenerate spectrum.\\

We have considered the possibility to 
determine the mass matrix in future  neutrino experiments. 
We find that the matrix can be reconstructed completely 
in the 
case of inverted mass hierarchy, provided that (i) the sensitivity 
to the absolute mass scale will reach $\sqrt{\Delta m_{atm}^2}$, 
(ii) the neutrinoless double beta decay will be discovered, 
(iii) stronger upper bounds on $s_{13}$ will be obtained. 

In the case of inverted ordering or degenerate spectrum 
the phase $\sigma$ becomes important. This phase cannot be determined, 
thus leaving large uncertainty in the structure of the 
$\mu\tau$-block.


\appendix
\section*{Appendix : general formulae}
\setcounter{equation}{0}
\renewcommand{\theequation}{A.\arabic{equation}}
\setcounter{subsection}{0}
\renewcommand{\thesubsection}{A.\arabic{subsection}}

We present explicit analytical expressions for the matrix elements.

The smallness of the parameter $s_{13}$ is very important for the analysis
of the matrix elements.
Defining
\beq
\begin{array}{l}
X\equiv xe^{i\phi_x} \equiv s_{12}^2 k e^{-2i\rho} + c_{12}^2\;,\\
Y\equiv ye^{i\phi_y} \equiv s_{12}c_{12}(1 - k e^{-2i\rho})\;,\\
Z\equiv ze^{i\phi_z} \equiv c_{12}^2 k e^{-2i \rho} + s_{12}^2\;,
\end{array}
\label{XYZ}
\end{equation} 
one can write the elements 
as series of powers of $s_{13}$:
\beq
\begin{array}{l}
{M}_{ee}/m_2 =Z-s_{13}^2 Z'\;,
\\
{M}_{e\mu}/m_2 =c_{13}(c_{23} Y -
s_{13}s_{23}e^{-i\delta} Z')\;,
\\
{M}_{e\tau}/m_2 = c_{13}(-s_{23} Y -
s_{13}c_{23}e^{-i\delta} Z')\;,
\\
{M}_{\mu\mu}/m_2 = c_{23}^2 X  + s_{23}^2 r e^{-2i\sigma}
-s_{13}\sin{2\theta_{23}} e^{-i\delta}Y
+ s_{13}^2 s_{23}^2 e^{-2i\delta} Z'\;,
\\
{M}_{\tau\tau}/m_2 = s_{23}^2 X  + c_{23}^2 r e^{-2i\sigma}
+s_{13}\sin{2\theta_{23}}  e^{-i\delta}Y
+s_{13}^2 c_{23}^2 e^{-2i\delta} Z'\;,
\\
{M}_{\mu\tau}/m_2 = s_{23} c_{23} (- X + r e^{-2i\sigma})
-s_{13}\cos{2\theta_{23}} e^{-i\delta}Y
-s_{13}^2 s_{23}c_{23} e^{-2i\delta} Z'\;,
\end{array}
\label{big}
\end{equation}
where
\beq
Z'\equiv z'e^{i\phi_{z'}} \equiv Z - r e^{2i(\delta-\sigma)}\;.
\label{zprime}
\end{equation}

We are interested in the limit $k\rightarrow 1$. For $k=1$, 
it follows from Eq.(\ref{XYZ}) that 
\beq
x=z=\sqrt{1-y^2}\;,
\label{xyz2}
\end{equation}
where $x$ is given in Eq.(\ref{xfx}).
Taking into account that $c_{12}>s_{12}$, it is easy to compute also 
the phases:
\beq
-\phi_x = \phi_z + 2\rho, ~~~~~~~
\phi_y=\frac{\pi}{2}-\rho\;,
\end{equation}
where $\phi_x$ is given in Eq. (\ref{xfx}). \\

Let us write explicitly the matrix $m^s$,
introduced in Eq.(\ref{0+s13}).
Defining
$$
\varphi_1 = \phi_{z'}-\phi_y-\delta\;,\qquad 
\varphi_2 = \phi_y-\delta\;,\qquad
\varphi_{\alpha\beta} = \arg M_{\alpha\beta}^0\;,
$$
we get, using Eq.(\ref{big}),
\beq
m^s= m_2\left(
\begin{array}{ccc}
0 & -s_{23}z'\cos{\varphi_1} & c_{23} z' \cos{\varphi_1}\\
\dots & -\sin{2\theta_{23}}\sqrt{1-x^2}\cos(\varphi_2-\varphi_{\mu\mu}) &
-\cos{2\theta_{23}}\sqrt{1-x^2}\cos(\varphi_2-\varphi_{\mu\tau})\\
\dots & \dots & 
\sin{2\theta_{23}}\sqrt{1-x^2}\cos(\varphi_2-\varphi_{\tau\tau})
\end{array}
\right)\;.
\label{os13}
\end{equation}
The maximal values of these corrections can be easily computed:  
using Eqs.(\ref{zprime},\ref{xyz2}), one finds
$$
{z'}^2=x^2+r^2+2rx\cos(2\delta-2\sigma-\phi_z) ~~\in~~ [(x-r)^2,(x+r)^2]\;.
$$

Finally, we give the explicit expression of the matrix 
$m^{\epsilon}$,
introduced in Eq.(\ref{0+s13}):
\beq
m^{\epsilon}=m_2\left(
\begin{array}{ccc}
-c_{12}^2 \cos(2\rho+\phi_z) & 
c_{23}c_{12}s_{12}\cos(2\rho+\phi_y) & 
s_{23}c_{12}s_{12}\cos(2\rho+\phi_y)\\
\dots & -c_{23}^2s_{12}^2\cos(2\rho+ \varphi_{\mu\mu}) &
c_{23}s_{23}s_{12}^2\cos(2\rho + \varphi_{\mu\tau})\\
\dots & \dots & 
-s_{23}^2s_{12}^2\cos(2\rho+\varphi_{\tau\tau})
\end{array}
\right)\;.
\label{oe}
\end{equation}
The cosines in Eq.(\ref{oe}) take always the values
$\pm 1$ for $\rho,\sigma = 0,\pi/2$. We have seen that
very small matrix elements usually appear for 
$\rho,\sigma \approx 0,\pi/2,\pi$ (see section \ref{hiera}).



\end{document}